\documentclass[11pt]{article}
\usepackage[utf8]{inputenc}
\usepackage{xcolor}
\usepackage{graphicx}
\usepackage[left=2.0cm, right=2.0cm, top=2.0cm, bottom=2.0cm]{geometry}
\usepackage{apacite}
\usepackage{booktabs}
\usepackage{amsmath}
\usepackage{caption}
\usepackage{float}
\usepackage{enumitem}
\usepackage{tabularx}

\AtBeginDocument{}

\title{\large \textbf{AUDIENCE CAPTURE, SELECTIVE EXPOSURE, AFFECTIVE ASSIMILATION OR IDEOLOGICAL SORTING? POLARISATION OF\\ CLIMATE POLITICS UNDER LOW MEDIA-PARTY PARALLELISM}}
\author{\normalsize Arttu Malkamäki\footnotemark[1] \footnotemark[2] \footnotemark[3] \and \normalsize Antti Gronow\footnotemark[1]}
\date{}

\begin{document}

\maketitle 
\footnotetext[1]{\footnotesize Faculty of Social Sciences, University of Helsinki, Finland}
\footnotetext[2]{\footnotesize Department of Computer Science, Aalto University, Finland}
\footnotetext[3]{\footnotesize Department of Communication, Stanford University, United States of America (Visiting scholar)}

{\let\abstractname\relax 
\begin{abstract}
\noindent Studies often attribute the polarisation of climate politics in majoritarian democracies to differing treatment of climate change by conservative and liberal media. How then does such polarisation occur in a system, where mainstream media do not systematically align with party positions? Focusing on how users in different ideological blocs posted and commented on news articles on Finnish Twitter between 2015 and 2023, we address four mechanisms: \textit{audience capture}, where outlets adapt to increasingly differentiated audiences, reinforcing their divergence through feedback between audience engagement and editorial incentives; \textit{selective exposure}, where users post ideologically congruent articles from the same outlets; \textit{affective assimilation}, where users communicate identical content with ideology-consistent valence; and \textit{ideological sorting}, where identity-driven alignment structures news-posting behaviour across multiple issues. Bayesian regression models indicate increasing divergence in outlet choice, with conservative-leaning users posting more tabloid content, in line with audience capture, and partial support for selective exposure, as within-outlet article posting becomes increasingly differentiated. Valence divergence increases only slightly over time, providing moderate support for affective assimilation, while issue-specific communities show weak alignment with blocs despite convergence over time. Our work refines theories of media effects beyond majoritarian democracies and contributes to understanding the current impasse in climate politics. \\
\\
\noindent \textit{hybrid media system}; \textit{mediatisation}; \textit{network analysis}; \textit{news media}; \textit{partisan sorting}; \textit{social media}
\end{abstract}}

\section*{\large INTRODUCTION}

Addressing climate change has become one of the most salient and contested issues in contemporary politics. Its increasing polarisation is commonly attributed to differences in how conservative and liberal news outlets cover the issue, particularly in media systems characterised by strong media-party parallelism \cite{roberts_climate_2025}. Media-party parallelism describes the extent to which news organisations align their editorial positions and audiences with political parties \cite{van_kempen_media-party_2007}. In contexts such as the United States, where partisan media have become deeply embedded in political competition, differences in news coverage have been shown to reinforce ideological divisions over climate change \cite{feldman_climate_2012, chinn_politicization_2020, mcallister_balance_2021}.

This explanation is less convincing in media systems characterised by low media-party parallelism. Finland represents such a case. A multi-party representative democracy, its news market is comparatively small, mainstream outlets are generally regarded as politically neutral, public trust in news remains high and editorial positions do not systematically align with political parties or their electorates \cite{newman_digital_2024}. Climate politics has nonetheless become increasingly polarised alongside the growing electoral success of the Finns Party, whose political agenda has become progressively more sceptical of environmental policies over the past decade \cite{lonnqvist_polarization_2020}. This raises a fundamental question. Why has climate politics become increasingly polarised in a country characterised by low media-party parallelism?

We address this question by examining how individual users belonging to different ideological blocs engage with climate news on Finnish Twitter between 2015 and 2023. Rather than assuming that polarisation emerges primarily from partisan media, we evaluate four competing mechanisms. First, \textit{audience capture} proposes that news organisations exploit digital affordances and growing political polarisation by tailoring and promoting content to increasingly distinct ideological audiences in order to maximise engagement \cite{chadwick_tabloids_2018, jurg_alex_2025}. Second, \textit{selective exposure} suggests that users preferentially share ideologically congruent articles, even when these originate from the same news outlets \cite{stroud_media_2008, tyler_partisan_2022}. Third, \textit{affective assimilation} proposes that users increasingly share identical articles with valence that is consistent with their prior ideological beliefs \cite{wilson_preferences_1989, lord_biased_2009}. Lastly, \textit{ideological sorting} suggests that alignment with a broader ideological identity increasingly structures patterns of news sharing across political issues, producing greater correspondence between ideological blocs and overall communication behaviour, analogous to partisan sorting in two-party systems \cite{baldassarri_partisans_2008}.

To evaluate these mechanisms, we draw on large-scale Twitter data, which provide a unique opportunity to observe news sharing and affective responses in a natural setting over time. As online platforms combine public interaction, user-generated content and algorithmically mediated information flows, Twitter data allow us to examine both the selection of political information and the ways in which it is communicated. This makes them particularly well suited to investigating the behavioural signatures of the mechanisms that may contribute to increasing political polarisation despite seemingly low media-party parallelism.

The remainder of the article is structured as follows. Second section develops the theoretical framework and derives hypotheses for each of the four mechanisms. Third section describes the data collection and pre-processing procedures. Fourth section details the operationalisations of the mechanisms. Fifth section reports the empirical results. Section 6 discusses the findings and their theoretical, methodological and practical implications, and concludes. 

\section*{\large THEORETICAL FRAMEWORK}

\subsection*{Polarisation beyond media-party parallelism}

Climate change has become one of the most polarising issues in contemporary politics, with ideological divisions increasingly structuring how citizens approach the underlying collective action dilemma \cite{aklin_prisoners_2020, falkenberg_growing_2022, judge_environmental_2023}. Comparative research on media systems often distinguishes countries by the extent to which their news outlets are institutionally and ideologically tied to political parties \cite{seymour-ure_political_1974, hallin_comparing_2004}. Where such media-party parallelism is strong, citizens arguably infer a great deal about the political content of a news story simply from knowing its source, and differential coverage of issues across outlets might itself generate or reinforce ideological divergence in audiences \cite{van_kempen_media-party_2007, martin_bias_2017}. The bulk of research on media-driven polarisation of climate politics builds on this logic, documenting how conservative outlets, especially in the widely studied anglophone systems, systematically downplay the scientific consensus on climate change while left-leaning outlets do the opposite or strive to further contextualise climate coverage \cite{feldman_climate_2012, ophir_politicization_2024, oneill_image_2013, bruggemann_beyond_2017, chinn_politicization_2020, mcallister_balance_2021}.

Much like Nordic media systems more generally, Finland, however, is an outlier. Its democratic corporatist media system has consistently been characterised by much lower parallelism, high professional autonomy among journalists and a strong, trusted public broadcaster \cite{van_kempen_media-party_2007, newman_digital_2024}. Editorial choices are, in relative terms, only loosely coupled to party politics, and audiences of the national mainstream and regional outlets tend to be ideologically heterogeneous. Yet climate change has clearly become a polarising issue in Finnish politics over the past decade, with the populist far-right Finns party adopting an increasingly obstructionist stance and issue positions on immigration and the environment becoming more tightly bundled together \cite{lonnqvist_polarization_2020, chen_polarization_2021}. If differential media treatment cannot straightforwardly explain this pattern, other mechanisms operating downstream of publication, at the point where users encounter, select and respond to news, become correspondingly more important to consider.

We distinguish four mechanisms through which ideological divergence in climate-related news engagement could emerge even without high media-party parallelism. These occur at different levels, namely media level (H1), individual level (H2 and H3) and system level (H4).

\subsection*{Media level: audience capture}

Our first mechanism concentrates on the concept of \textit{audience capture}, whereby media actors adjust their coverage and promotional strategies in response to the preferences of an increasingly vocal and demanding segment of their audience \cite{jurg_alex_2025}. As outlets compete for attention under conditions of audience fragmentation and platform-mediated distribution, they have an incentive to differentiate their offering by tailoring and promoting content that appeals disproportionately to particular ideological audiences \cite{chadwick_tabloids_2018, gentzkow_ideological_2010}. This does not necessitate any formal media-party alignment. As \citeauthor{chadwick_tabloids_2018} (2018) argue, tabloid outlets in particular may simply discover, for example through the feedback loops built into most online platforms \cite{zeitzoff_how_2017, frimer_incivility_2023}, that a particular style of climate coverage travels further among certain audiences and adjust its editorial or promotional emphasis accordingly. Journalistic incentives might well contribute to this process \cite{langer_political_2021}. Under audience capture, ideological clusters should increasingly diverge in which types of outlet they draw upon for climate news, even while formal editorial neutrality is maintained. We set out our initial hypothesis:

\begin{itemize}
    \item[H1] \textit{Ideological divergence in outlet choice drawn upon \\ for climate-related content increases over time}. 
\end{itemize}

\subsection*{Individual level: selective exposure and affective assimilation}

The second mechanism is \textit{selective exposure}. Classic accounts describe how citizens gravitate towards information that confirms their prior beliefs and avoid information that challenges them, a pattern documented extensively in online news environments and beyond \cite{stroud_media_2008, barbera_tweeting_2015, tyler_partisan_2022, mangold_ideological_2024}. Even where all ideological groups are formally exposed to the same outlets, users may selectively pick out and share only those specific articles that confirm their prior beliefs, leaving other coverage from the same source unshared. In our setting, we therefore conceptualise selective exposure at the level of individual articles rather than news outlets. It would therefore manifest as systematic ideological divergence in which articles from otherwise common outlets are selected and shared. Our second hypothesis reads as follows:

\begin{itemize}
    \item[H2] \textit{Conditional on news outlet, ideological blocs will increasingly \\ select different climate-related articles from the same outlets}. 
\end{itemize}

Third, we consider a mechanism which we call \textit{affective assimilation}. It builds on theories of politically motivated reasoning in general and biased assimilation in particular, which suggest that individuals tend to evaluate political information through the lens of their prior preferences, expectations and beliefs \cite{taber_motivated_2009, lord_biased_2009, bisgaard_bias_2015, gronow_external_2025}. We however use affective assimilation to refer specifically to the affective expression of biased assimilation, conceptualising it as the tendency for the emotional appraisal and communication of political information to conform to prior ideological commitments \cite{han_between_2026}. This reflects but also deviates from the affective expectation model commonly used to study affective assimilation \cite{wilson_preferences_1989}, as our focus is on alignment with ideological orientation rather than with a specific prior expectation. This distinction is useful because interpretive frames are difficult to observe directly in naturally occurring communication, whereas differences in affective valence provide a more tractable signal. Applied to our setting, even where the same article from the same outlet is shared across ideological blocs, users should frame that content in consistently different emotional terms depending on their ideology. The corresponding hypothesis is the following:

\begin{itemize}
    \item[H3] \textit{Conditional on article, ideological blocs will increasingly express different \\ valence when posting the same climate-related articles over time}. 
\end{itemize}

\subsection*{System level: ideological sorting}

Our final, fourth mechanism, \textit{ideological sorting}, shifts attention from climate change in isolation to its place within a broader identity-driven structure. Research on partisan sorting in two-party systems such as the US has shown that attitudes towards seemingly unrelated political issues, such as immigration, security or economic redistribution, increasingly align along a common dimension characterised by strong partisan identity \cite{baldassarri_partisans_2008, iyengar_affect_2012, mason_i_2015}. Because Finland's multi-party system is organised around broader ideological blocs rather than two major parties, we refer to this process as ideological sorting rather than partisan sorting. The underlying logic, however, is identical, and broader ideological blocs simply provide a more meaningful unit of analysis than individual parties in this context. As ideological identities become increasingly coherent, patterns of news sharing on climate change should increasingly resemble those observed for other salient political issues. Empirical evidence already points in this direction, with issue alignment documented between climate and immigration attitudes specifically within the Finnish context \cite{lonnqvist_polarization_2020, chen_polarization_2021}. If climate-related news engagement is simply one expression of broader ideological sorting, the communities emerging not only from climate news sharing but across multiple salient issues should increasingly correspond to users' underlying ideological affiliations. Climate polarisation would therefore be better understood as a manifestation of generalised ideological sorting than as an issue-specific phenomenon -- hence our last hypothesis:

\begin{itemize}
    \item[H4] \textit{Alignment between users' ideological affiliations and the structures emerging \\ from news sharing patterns will increase over time across political issues}. 
\end{itemize}

These four mechanisms are not mutually exclusive and may operate simultaneously, yet each is bound to leave a distinct trace in observational data, which we also exploit in our operationalisation strategy. They represent complementary explanations for increasing climate polarisation under conditions of low media-party parallelism. Audience capture attributes polarisation to media organisations, selective exposure to individual information selection, affective assimilation to ideological differences in emotional communication and ideological sorting to wider identity-driven alignment across political issues.

Each hypothesis therefore targets a different point in the chain running from publication to reception, which allows us to adjudicate between explanations that are frequently conflated in the literature on media and polarisation. When considered jointly, strong support for H1 and weak support for H2 would suggest that polarisation in a low-parallelism system operates essentially through the types of outlets on offer rather than through fine-grained curation within outlets, while support for H3 would indicate that a meaningful share of polarisation is generated after exposure, not through interpretation but through selection. Support for H4 would in turn suggest that whatever polarisation is observed on climate news is not issue-specific but a downstream consequence of more general ideological sorting occurring within the Finnish political system.

\section*{\large DATA}

To address our hypotheses, we focussed on a very specific act, namely the posting of news articles on an online platform. Consider, for example, the situation where a politician pastes a hyperlink to an online post that leads to the website of a news outlet, with some commentary such as ``A great perspective on the ongoing erosion of social norms in politics.''. To study such behaviour over time and at scale, we resorted to observational digital trace data from the Twitter. At least prior to the acquisition and subsequent restructuring of the platform as \textit{X} on 23 July 2023, Twitter served as a major, algorithmically mediated arena for political communication among politicians, officials, journalists and politically active members of the public, also in Finland \cite{koivunen_trust_2022, molyneux_legitimating_2022}. Although the platform has changed substantially since then, these data remain relevant because they capture communication during a period when Twitter occupied a uniquely influential position in the political information ecosystem.

As our hypotheses arose in part from the peculiar media-political context of Finland, we began by collecting all posts in the Finnish language since the right-leaning Sipilä government took office in Finland (4 April 2015 to 14 April 2019) until the last day of the left-leaning Marin cabinet (15 April 2019 until 2 April 2023). In so doing, we first passed a climate-specific keyword query to Twitter's academic programming interface in June 2023, and then applied a more specific partial match keyword filter to the output, namely one without energy-specific keywords due to the 2022 Russian invasion of Ukraine and the resulting energy crisis in Europe (see Appendix). After extracting all posts with a hyperlink (i.e. omitting retweets, quote retweets and replies), there were 40,209 unique users in our data who had authored a total of 428,654 such posts over the roughly eight-year period.

The data underwent several pre-processing stages. The most laborious task was the valence detection, the technical details of which we have relegated to the Appendix. In a nutshell, the task entailed cleaning the text fields by removing hyperlinks, hashtags and recurring headline fragments, distinguishing between comments targeting the substantive content of the article and those addressing journalistic practices, and fine-tuning a large pre-trained Finnish language model to classify each post as negative, neutral or positive, followed by a post-classification adjustment in which low-confidence neutral predictions were reassigned to either negative or positive where the model exhibited a clear preference between the two.

In addition, assigning each user to an ideological bloc was a fundamental criterion for addressing our hypotheses. Again, we have relegated much of the technical detail to the Appendix, providing here only an overview of the procedure. We inferred users' leanings separately for the two parliamentary cycles. We constructed two graphs from separately collected retweets, commonly interpreted as indications of agreement, endorsement or trust \cite{metaxas_what_2015}, between users containing a reference to a Finnish political party. We then a fitted a planted partition model, a version of the stochastic block model, to each graph to identify relatively dense clusters of users \cite{zhang_statistical_2020}. The resulting three clusters in both cycles were validated against the observed positions of known parliamentary election candidates, again separately for 2019 and 2023 elections, and interpreted as wide ideological blocs, whether Conservative Right (CR), Liberal Left (LL) or Moderate Right (MR).

In the Finnish context, the resulting blocs corresponded to known party alignments \cite{isotalo_polarisoituuko_2020}, with the CR comprising candidates of parties such as the Finns Party and a fraction of the National Coalition Party, the LL including the Social Democratic Party, the Green League and the Left Alliance, and the MR aligning most closely with the Centre Party as well as the majority of National Coalition Party candidates. A noteworthy proportion of candidates in the CR bloc represented parties without prior parliamentary representation in Finland, many of which were positioned on the far right of the political spectrum. In the Appendix, we provide the weekly participation by bloc and polarisation between each bloc pair based on users' climate-related retweets. These trends match our understanding of the political shifts and allegiances in national climate politics over the study period, further validating the robustness of our bloc inference. Yet we also observe a marked increase in participation in climate politics on Finnish Twitter following Finland's hottest-ever July temperature in 2018, with participation remaining persistently elevated thereafter.

After harmonising the hyperlinks to enhance mapping between users and specific articles (e.g. similar hyperlinks leading to the same article), solving instances of the same user posting the same article multiple times (i.e. by retaining the one yielding the highest engagement metrics) and narrowing the data down to valid domains of 77 valid outlets, we omitted all (fragments of) posts addressing journalistic practices and all users for whom we were unable to assign an ideology, eventually leaving us with 10,646 users and 107,005 posts. Only 11\% of the remaining posts exhibited non-neutral valence, suggesting that emotional commentary on climate news was relatively uncommon on Finnish Twitter, although our classifier had been tuned to be conservative.

\begin{figure}[ht]
    \centering
    \captionsetup{justification=centering,font=small}
        \includegraphics[width=0.99\textwidth,trim={0pt 0pt 0pt 0pt},clip]{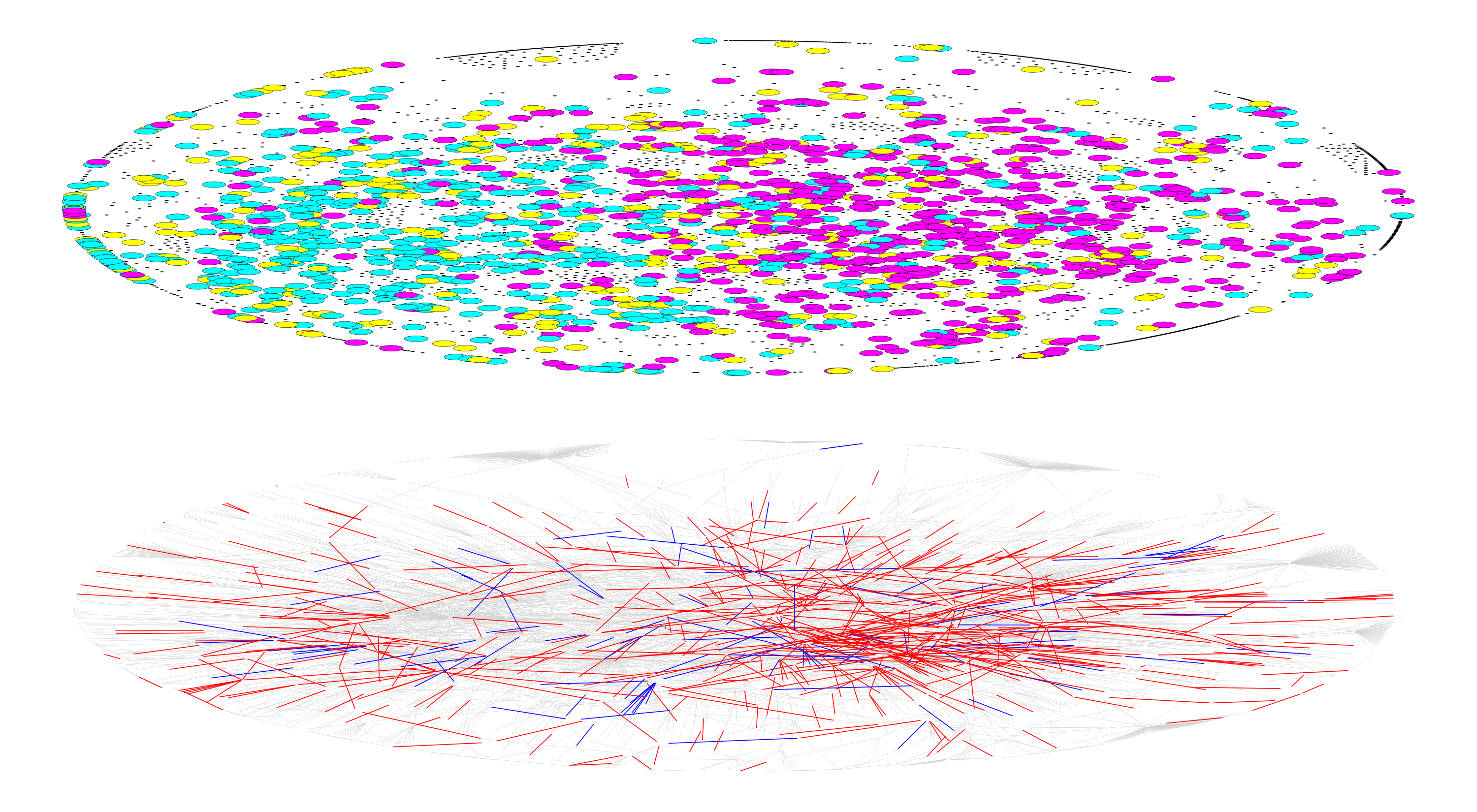}
        \caption{Bipartite graph representation of the data from July 15 to October 14, 2022; vertex colours indicate inferred ideological blocs of users (\textcolor[HTML]{ff00ff}{{\textbullet}} Conservative Right, \textcolor[HTML]{00ffff}{{\textbullet}} Liberal Left, \textcolor[HTML]{ffff00}{{\textbullet}} Moderate Right, \textcolor[HTML]{000000}{{\textbullet}} article-hyperlink); edge colours indicate inferred valence of the post (\textcolor[HTML]{ff0000}{{\textbf{\textemdash}}} Negative, \textcolor[HTML]{d3d3d3}{{\textbf{\textemdash}}} Neutral, \textcolor[HTML]{0000ff}{{\textbf{\textemdash}}} Positive).}
        \label{fig:bipartite_data_example}
\end{figure}

Eventually, we had 5,236 users and 35,428 posts with a hyperlink to a climate news left in our data for the Sipilä cycle, and 8,267 users and 71,577 posts for the Marin cycle, all with the necessary meta-data concerning user leaning and post sentiment. As observing polarisation by definition comes down to observing a process of divergence, a key criterion for testing our hypotheses was time. Hence, and given that our data afforded us to integrate time in our analyses, we further divided the entire eight-year period into 32 three-month slices to consider and control for the effects of major events and subsequent peaks of user activity. A fixed interval was necessary to keep track of the fluctuations in activity, while an alternative approach, such as one using an adaptive window of a particular size (e.g. 1,000 posts) ought to have been more appropriate if our interest was on a particular event or discussion becoming viral and unfolding very rapidly, in the course of hours and days (e.g. \citeauthor{xia_russian_2024}, 2024). Data aggregation also ensured that we had enough information to follow through our analyses in a robust manner. Figure~\ref{fig:bipartite_data_example} visualises one such slice, from July 15 to October 14, 2022, as a bipartite graph in which vertices represent users who connect to one another through an article if they posted, and potentially commented on, the same article in the same period at least once. Vertex colours in the ``ceiling'' map to the inferred blocs of users and edge colours in the ``floor'' to the inferred valence of posts. The graph layout algorithm places vertices with more similar connectivity patterns closer to one another.

Using observational-longitudinal Twitter data carries several advantages over alternative approaches to address our hypotheses, yet certain caveats warrant attention. The retrospective extraction of data for instance meant that we were unable to recover any content that had already been taken down from the server, for whichever reason (e.g. self-removal or rule violation), the extent of which we however consider minimal and as causing a trivial pattern of missingness (see also \citeauthor{pfeffer_this_2023}, 2023). An implication of Twitter's acquisition in July 2023, after which the algorithm changed and some users left the platform, is that our empirical data does not necessarily generalise directly to the current media landscape in Finland or elsewhere, yet we still consider the theoretical mechanisms underlying our hypotheses highly generalisable. Even with Twitter alone, we however caution against drawing far-reaching conclusions based on the (partially) algorithmically confounded behaviour of (presumably) politically most active segment of the Finnish population. At the time of analysis, appropriately validated large language models (LLMs) for the Finnish language were not available. Considering LLMs for similar tasks in the future seems promising however and definitely warrants more attention in improving valence inference.

\section*{\large OPERATIONALISATIONS}

Instead of modelling polarisation, a multi-dimensional macro-level phenomenon, directly, we operationalise the four mechanisms using measures that correspond to distinct stages of the news-posting process: outlet choice (H1), article selection within outlets (H2), affective responses to common content (H3) and the broader organisation of issue-specific news-posting patterns (H4). We estimate the main associations using Bayesian regression models implemented in \textit{brms} \cite{kruschke_bayesian_2018, burkner_brms_2017}. We however use the default prior specification rather than imposing substantive researcher-specified priors, allowing the data to determine the posterior estimates while retaining the regularisation built into the modelling framework where applicable. Bayesian estimation is particularly useful for our analyses because it accommodates the hierarchical structure of the data, propagates uncertainty through complex model-based predictions and contrasts, and provides directly interpretable posterior distributions for both parameters and the predicted quantities relevant to our hypotheses. This also allows uncertainty in these derived quantities to be assessed directly through posterior credible intervals.

For substantive interpretation, we report posterior means and 95\% credible intervals, using estimated marginal means and conditional effects to express the fitted models as predicted quantities corresponding directly to our hypotheses. Posterior contrasts assess differences between ideological blocs and, where relevant, between their temporal trajectories \cite{searle_population_1980}, while other model terms are interpreted using their posterior coefficients. Model adequacy is assessed using Bayesian $R^2$, $\widehat R$ convergence diagnostics, effective sample sizes and posterior predictive checks of distributional shape, empirical cumulative distribution function, mean, standard deviation, minimum and maximum values, and average predicted values, alongside traditional residual-versus-predicted-value plots \cite{gelman_r-squared_2019, gabry_visualization_2019, vehtari_rank-normalization_2021}. We provide full fixed-effect estimates for our main models as online supplementary material. 

To test H1, we examine whether ideological blocs differ systematically in their propensity to post material from different categories of news outlets and whether these differences change over time. We do not measure the share of all posts directed towards each outlet category but the proportion of users in each ideological bloc who posted at least once from a given outlet category within a time window. This distinction is important because the hypothesis concerns the audiences that outlets appear to attract or target, rather than simply differences in the volume of posting generated by users with different levels of activity. For each three-month window, we therefore calculate the number of users in each ideological bloc who posted at least one climate-related article from each outlet category and relate this to the number of active users in the corresponding bloc. We estimate these proportions using a binomial distribution, retaining the number of users who post and the total number of participants respectively as the numerator and denominator. The models include ideological bloc, outlet category and their interaction, together with ideological bloc and time interactions. Increasing differences between blocs would be consistent with audience capture, particularly if the pattern is concentrated in outlet categories with stronger incentives to appeal to ideologically distinctive audiences.

To classify outlets in a way that supports both substantive interpretation and aggregation, we manually inspected the domains occurring in the data and considered their prominence and role within the Finnish media system \cite{media_audit_finland_kansallinen_2022, newman_digital_2024}. The final categorisation distinguishes financial, fringe (mainly nationalist), mainstream (widest circulation), regional, partisan, tabloid and miscellaneous (still potentially salient) outlets, with less salient outlets omitted from our analyses. See Appendix for the corresponding domain classification. Aggregating outlets into categories reduces the influence of individual domains and allows the analysis to capture broader differences in outlet type rather than idiosyncratic behaviour by particular organisations.

H2 concerns selective exposure within outlets. The key question is therefore not whether ideological blocs use different outlets, but whether users belonging to different blocs select different articles even when they obtain news from the same outlet. We operationalise this as within-outlet polarisation in the sharing of specific articles. For each domain and time window, we construct a binary user-article bipartite graph in which an edge indicates that a user shared a particular article from that domain. We then project this bipartite graph onto a weighted user-user graph, such that users are connected to the extent that they share the same articles. Because raw co-sharing counts are strongly affected by differences in user activity, we normalise projected edge weights using standard cosine similarity based on the overlap in the articles shared by two users relative to their respective activity levels \cite{leifeld_discourse_2017}.

We then calculate the adaptive external-internal index (AEI) for each pair of ideological blocs \cite{salloum_separating_2022}. The index extends the external-internal index of \citeauthor{krackhardt_informal_1988} (1988) by adjusting for potential differences in group size. We however generalise it ourselves to account for the sum of edge weights instead of mere edge counts. AEI ranges between -1 and 1, and captures the extent to which the weighted edges between users fall within rather than between ideological blocs. Values towards the positive end of the scale indicate greater separation between the blocs, whereas values towards the negative end indicate greater within-bloc cohesion. Values around zero indicate comparatively little separation between blocs after accounting for bloc composition. We also impose a minimum validity threshold of ten active users from each ideological bloc within a domain and time window. This prevents the index from being driven by very small numbers of users and ensures that the comparison between two blocs is based on a minimally viable set of observations. Increasing within-outlet AEI over time would provide evidence consistent with selective exposure -- that is, users would increasingly encounter and post different articles from the same outlet.

Our H2 model uses a Gaussian distribution to predict pairwise AEI over time while accounting for differences in outlet category, outlet and outlet popularity. Outlet category enters as a fixed effect to capture systematic differences in baseline within-outlet polarisation, while an outlet-level random intercept accounts for persistent differences between individual outlets. We also include the log-transformed geometric mean number of users from the two ideological blocs to control for differences in the scale of their activity.

H3 asks whether users from different ideological blocs respond differently to the same news content. We therefore measure affective divergence at the article level rather than at the outlet level because the relevant comparison here is between the emotional valence expressed by different ideological blocs when sharing an identical article. For every article with at least two users from both ideological blocs, we calculate the valence distribution separately for each bloc and compare the resulting distributions between each bloc pair using the Jensen-Shannon distance (JSD) measure, a symmetric and bounded measure of dissimilarity between probability distributions \cite{lin_divergence_1991}. A JSD of zero indicates identical valence distributions, whereas larger values until one indicate increasingly divergent affective responses.

Since many articles have no observable divergence between the blocs, we distinguish between two related outcomes. First, we model whether the JSD is non-zero via a Bernoulli distribution, capturing the probability that the two ideological blocs express any discernible difference in valence when sharing the same article. Second, conditional on a non-zero JSD, we model the magnitude of that divergence. This distinction separates the occurrence of affective divergence from its intensity and avoids treating a large number of articles with identical valence distributions as equivalent to articles where the blocs respond in substantially different ways. The resulting article-level JSD measures are then modelled over time and across ideological pairs via a beta distribution. The principal quantity of interest is whether affective divergence becomes more prevalent or more pronounced over time and whether this temporal pattern differs systematically between ideological blocs. As in H2, we similarly control for differences in outlet and article-level scale of activity, but omit outlet category because it does not belong to the hypothesised mechanism. 

Lastly, H4 examines whether the blocs inferred independently from users' broader political behaviour on Twitter are reflected in the ``communities'' that arise from issue-specific news-co-posting patterns. The hypothesis therefore concerns the correspondence between a fixed ideological classification and evolving communities inferred not only from climate but also from the immigration, inequality (i.e. redistribution of income and wealth) and security (i.e. foreign policy and national defence) news-posting patterns. The three additional issues were chosen as they represent distinct yet persistently salient areas of political contestation throughout the study period, while providing sufficient variation in issue content to assess whether the patterns observed for climate extend to other issues. Data for these issues were collected separately (see Appendix). For each issue and time slice, we first construct a user-article bipartite graph and then project it onto a user-user graph using the same projection logic as earlier under H2. We then infer communities from the resulting issue-specific graphs using a similar non-parametric block model selection approach as we did to infer the ideological blocs (see Appendix).

To assess the correspondence between issue-specific community structures and the fixed ideological blocs, we use the normalised version of the reduced mutual information (RMI) measure \cite{newman_improved_2020}. RMI provides an information-theoretic measure of alignment between two clustering solutions while correcting the standard measure for effects arising from the sizes and numbers of clusters. Values closer to one indicate greater correspondence. Because the measure is based on information expressed on a logarithmic scale, increases in RMI do not represent equal increments in the amount of information shared between partitions. This is particularly relevant for small graphs, where the number and size of clusters constrain the amount of mutual information and make RMI estimates sensitive to small changes in the underlying partition. RMI might also take values below zero when the correspondence is lower than expected under its adjustment for chance.

As we calculate RMI separately for climate, immigration, inequality and security over time, we distinguish issue-specific polarisation from a more general process of ideological sorting. If climate polarisation simply reflects a broader tendency for political identities to structure news-posting behaviour, the communities emerging from climate news should increasingly correspond to the same fixed ideological blocs that structure other issue-based news-co-posting communities. A weaker correspondence for other issues would suggest that the alignment observed for climate cannot be attributed solely to ideological sorting.

Across our analyses, despite all our observations coming from three-month slices, we model time as a categorical predictor aggregated into six-month periods. This allows us to estimate period-specific differences without imposing a particular functional form on temporal change and avoids assuming that changes between adjacent periods reflect a smooth underlying trend. It also allows the models to absorb period-specific shocks, such as parliamentary elections (2019, 2023), political events (e.g. COVID-19 in 2020 and Russian invasion of Ukraine in 2022) or major developments in climate politics (e.g. reports, pledges or actions). Next, we turn our attention to the results for our regression models.

\section*{\large RESULTS}

\subsection*{Audience capture}

\begin{figure}[ht]
    \centering
    \captionsetup{justification=centering,font=small}
        \includegraphics[width=0.99\textwidth,trim={0pt 0pt 0pt 0pt},clip]{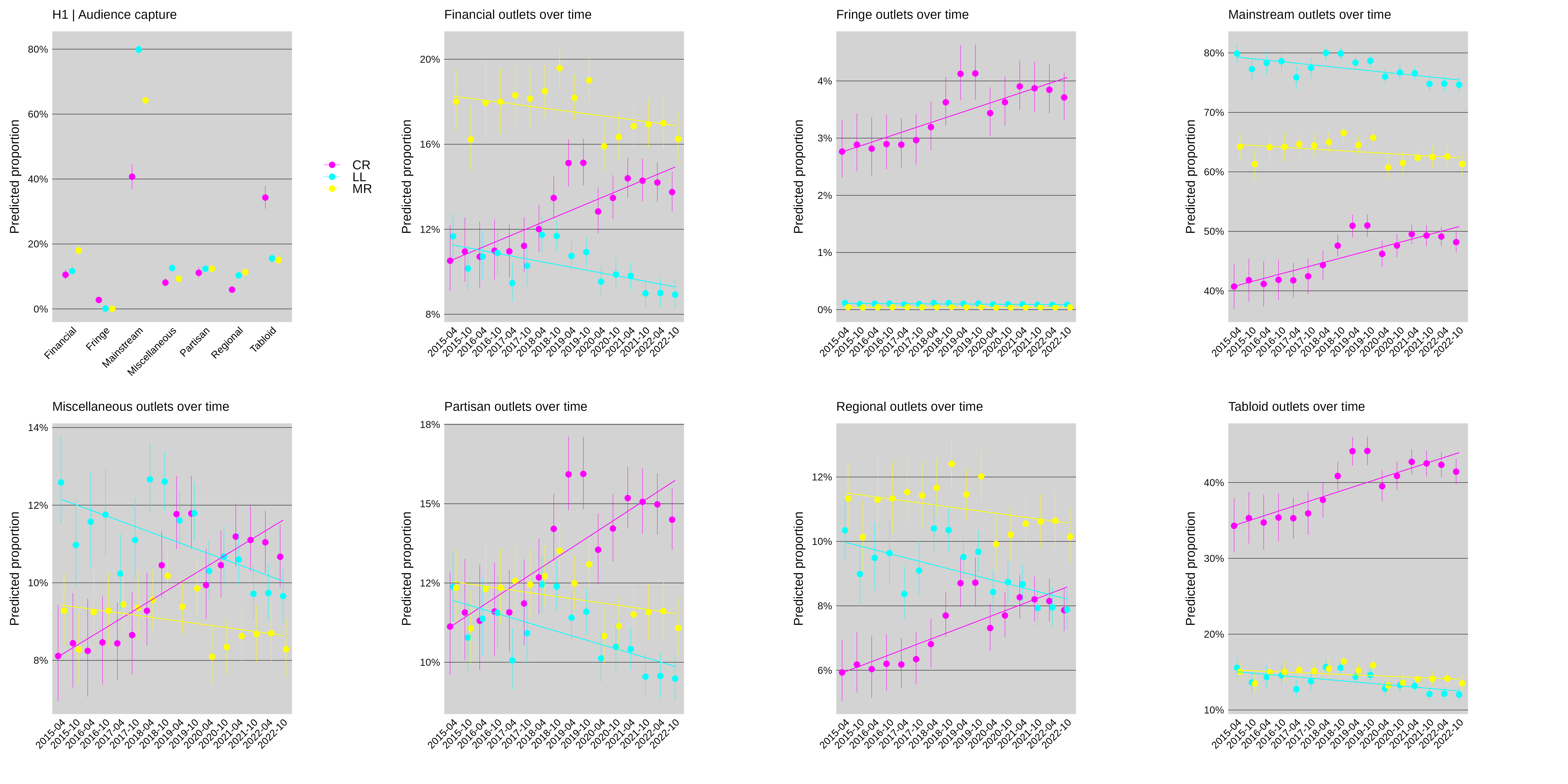}
        \caption{Predicted proportion of posting users belonging to each ideological bloc across outlet categories and over time; points denote posterior means and error bars indicate 95\% credible intervals; trend lines fitted separately to enhance interpretability.}
        \label{fig:audience_capture}
\end{figure}

Figure~\ref{fig:audience_capture} presents model-estimated probabilities of posting from each outlet category by ideological bloc and over time. The results show notable and persistent differences in outlet choice, alongside a gradual shift in the composition of engagement over the study period.


The clearest differentiation concerns mainstream and tabloid outlets. Mainstream outlets have the highest predicted participation among all three blocs, but the probability is substantially higher for LL users ($0.77$, $[0.77,\,0.78]$) than for MR ($0.64$, $[0.63,\,0.64]$) or CR users ($0.46$, $[0.45,\,0.47]$). This ordering is reversed for tabloids, where CR users have a predicted probability of $0.39$ $[0.38,\,0.40]$, compared with $0.14$--$0.15$ for LL and MR. The corresponding odds ratios indicate that CR users are approximately four times as likely as LL users ($OR=4.04$, $[3.80,\,4.28]$) and 3.75 times as likely as MR users ($OR=3.75$, $[3.51,\,3.99]$) to post tabloid content, while for mainstream outlets the direction reverses, with CR users much less likely than LL ($OR=0.25$, $[0.23,\,0.26]$) or MR users ($OR=0.49$, $[0.46,\,0.51]$) to post mainstream content.

Financial and regional outlets show weaker but still visible differentiation, with MR users most likely to post from both categories ($0.18$ $[0.17,\,0.18]$ and $0.11$ $[0.10,\,0.12]$ respectively). CR and LL users trail MR by comparable margins in both categories ($OR$s ranging $0.54$--$0.78$). Miscellaneous and partisan outlets are more similar across blocs. Fringe outlets are rarely posted overall, with a predicted probability that is small but non-zero for CR users ($0.03$, $[0.03,\,0.04]$) and effectively zero for LL and MR.

Over time, the same broad structure holds, though CR participation in mainstream outlets rises somewhat from 2015 ($0.41$) to 2022 ($0.48$), narrowing the mainstream gap as climate politics becomes more popular among CR users. Tabloid outlets show the opposite pattern and the clearest evidence of CR concentration, with CR tabloid posting rising from $0.34$ in 2015-04 to $0.44$ around the 2019 election (widely dubbed ``the climate election'' following an intense period of international issue attention) and remains elevated at $0.41$ by 2022-10, well above LL and MR, which stay around $0.12$--$0.16$ throughout. This identifies a persistent increase in CR tabloid engagement from the late 2010s onwards.

Financial outlets show a more moderate but substantively important shift. MR users remain most likely to post financial content throughout, but CR users rise from near-parity with LL in 2015 ($0.11$ vs.\ $0.12$) to a higher predicted probability than LL by 2022 ($0.14$ vs.\ $0.09$), possibly reflecting heightened attention to the fiscal consequences of COVID-19 and the run-up to the 2023 election.

The model provides an exceptionally strong account of the observed outcome distribution (Bayesian $R^2=0.97$, $[0.97,\,0.97]$), with satisfactory convergence ($\widehat R=1.00$ throughout) and posterior predictive checks capturing the observed mean, dispersion and range well.

Even though our data identify differentiated audience engagement rather than directly demonstrating editorial adaptation, these findings are consistent with audience capture. Ostensibly non-partisan news organisations appear to exploit online affordances to tailor content to increasingly polarised audiences.

\subsection*{Selective exposure}

\begin{figure}[ht]
    \centering
    \captionsetup{justification=centering,font=small}
        \includegraphics[width=0.99\textwidth,trim={0pt 0pt 0pt 0pt},clip]{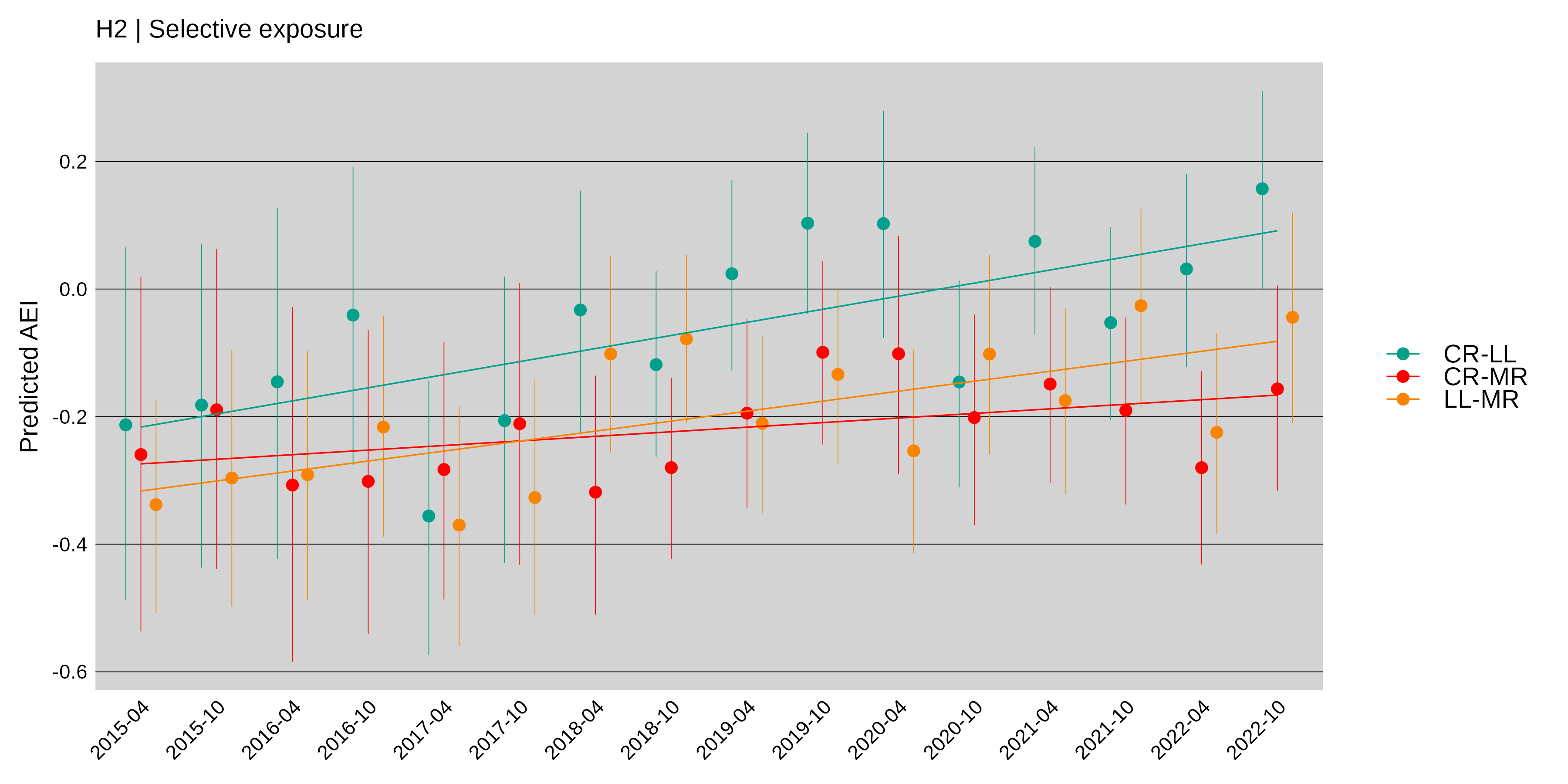}
        \caption{Predicted within-outlet polarisation, measured using the adaptive external-internal index (AEI), between pairs of ideological blocs over time; points denote posterior means and error bars indicate 95\% credible intervals; trend lines fitted separately to enhance interpretability.}
        \label{fig:selective_exposure}
\end{figure}

In Figure~\ref{fig:selective_exposure} we present the model-estimated AEI for each pair of ideological blocs over time, where AEI ranges from $-1$ to $1$. Negative values indicate that co-posted articles from the same outlet are more often shared between blocs, positive values indicate that they are more often shared within blocs. The results show some movement towards greater within-outlet segregation, albeit concentrating mainly in the CR-LL relationship.


In the early period, all three bloc pairs have negative estimated AEIs, indicating that users more often co-post the same articles across blocs than within blocs when those articles come from the same outlet. In 2015-04, the predicted AEI is $-0.21$ $[-0.49,\,0.07]$ for CR-LL, $-0.26$ $[-0.54,\,0.02]$ for CR-MR and $-0.34$ $[-0.51,\,-0.17]$ for LL-MR, the clearest early evidence of negative within-outlet AEI, with CR-LL and CR-MR credible intervals still including zero.

Over time, CR-MR and LL-MR remain mostly negative, with several periods credibly below zero for each pair (e.g. CR-MR: $-0.28$ to $-0.32$ in 2018; LL-MR: $-0.34$ to $-0.37$ in 2015--2017), while both pairs edge closer to zero in later waves without crossing it. CR-LL, in contrast, shows the clearest upward movement, rising from $-0.21$ $[-0.49,\,0.07]$ in 2015-04 to positive estimates in several later periods, including 2019-10 ($0.10$, $[-0.04,\,0.25]$) and 2022-10 ($0.16$, $[0.00,\,0.31]$). This movement into positive territory around 2019--2020 coincides with the 2019 election and the ensuing change in government that yielded an ambitious national climate pledge. The overall temporal pattern therefore does not indicate a general move towards within-outlet segregation across all ideological blocs, as CR-MR remains persistently negative, LL-MR weakens towards parity in some later periods and CR-LL is the only pair that approaches positive values. AEIs around zero, however, mean that selection does occur, with users posting both the same and different articles from the same outlet roughly to an equal extent.

The pairwise posterior contrasts reinforce this interpretation. In the early waves there is little evidence that the three pairwise AEIs differ from one another, with all contrast intervals including zero through 2017-10. From 2018 onwards, CR-LL becomes credibly more differentiated than both CR-MR and LL-MR in several periods (contrasts of roughly $0.20$--$0.36$), whereas differences between CR-MR and LL-MR are generally uncertain, with only 2018 showing CR-MR credibly more negative than LL-MR. These contrasts indicate that the main separation is between CR-LL rather than a three-way increase in within-outlet segregation across ideological blocs.

The control variables provide some evidence that activity is related to AEI. The coefficient for log-transformed geometric mean of participation across the two blocs is negative ($-0.07$, $[-0.12,\,-0.02]$), indicating that more active bloc-pair observations tend to be more negative on the AEI scale (i.e., co-posting is more likely to occur between blocs than within blocs for articles from the same outlet). None of the outlet-category coefficients differs credibly from the financial reference category, providing little evidence that baseline within-outlet AEI varies systematically across outlet categories after accounting for domain-level heterogeneity and activity.

The model captures the variation in AEI modestly, with a Bayesian $R^2$ of $0.24$ $[0.19,\,0.28]$. Convergence and sampling diagnostics are satisfactory, with $\widehat R=1.00$ for all parameters and adequate effective sample sizes across the model. Posterior predictive checks indicate that the model broadly reproduces the observed distribution of the outcome.

The findings are consistent with selective exposure operating asymmetrically across the ideological space, especially between CR and LL users. Yet there is scant evidence for a general pattern of within-outlet selection of news dominating.

\subsection*{Affective assimilation}

\begin{figure}[H]
    \centering
    \captionsetup{justification=centering,font=small}
        \includegraphics[width=0.99\textwidth, trim={0pt 0pt 0pt 0pt}, clip]{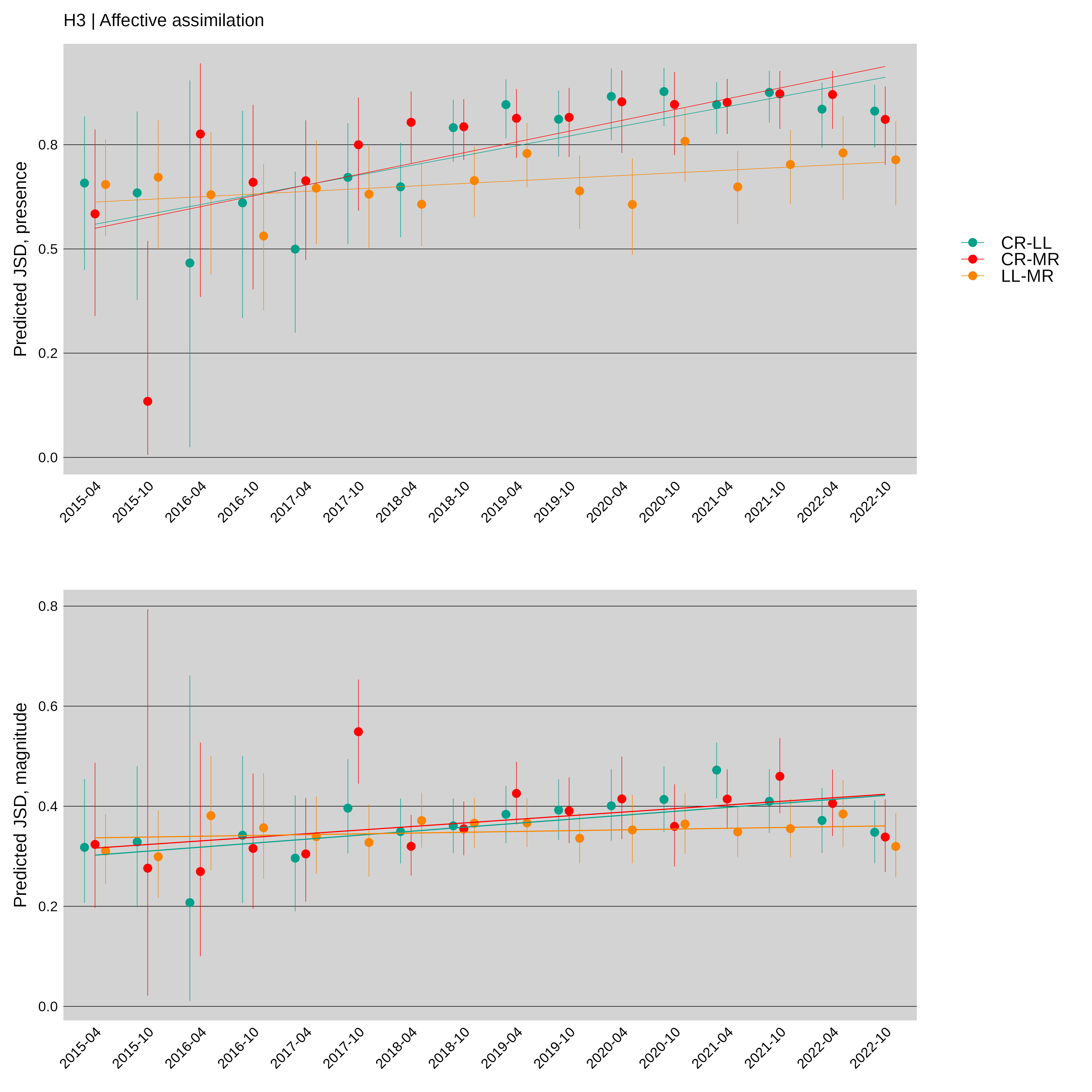}
        \caption{Predicted presence (upper) and magnitude (lower) of affective divergence, measured using Jensen-Shannon distance (JSD), between pairs of ideological blocs over time; points denote posterior means and error bars indicate 95\% credible intervals; trend lines fitted separately to enhance interpretability.}
        \label{fig:affective_assimilation}
\end{figure}

As regards our H3 focusing on whether ideological blocs respond with a systematically different valence to identical climate-related articles. Figure~\ref{fig:affective_assimilation} presents the predicted presence of non-zero JSD in the upper panel and the predicted magnitude of JSD conditional on divergence being present in the lower panel, showing that divergence becomes more likely over time, while its conditional magnitude remains moderate and only weakly differentiated across bloc pairs.


The first model estimates whether any affective divergence occurs. The predicted probability of non-zero JSD is already substantial at the beginning of the observation period and generally increases thereafter. For CR-LL, the probability rises from $0.66$ $[0.45,\,0.82]$ in 2015-04 to $0.85$ $[0.77,\,0.91]$ in 2019-04, remaining high at $0.83$ $[0.74,\,0.89]$ in 2022-10. CR-MR follows a similar trajectory, from $0.58$ $[0.34,\,0.79]$ to $0.81$ $[0.72,\,0.88]$ and $0.81$ $[0.70,\,0.89]$. LL-MR is also often above one half but is lower and more stable, moving from $0.65$ $[0.53,\,0.76]$ to $0.73$ $[0.65,\,0.80]$ and $0.71$ $[0.60,\,0.81]$. Thus, affective divergence is not rare as such. Instead, the main temporal pattern is that the presence of non-zero divergence becomes especially likely for pairs involving CR from the late 2010s onwards.

However, the presence of divergence should be distinguished from its size. Of the 4,062 pair-wise JSD observations, 41\% show zero divergence, leaving 2,382 observations for the conditional magnitude model. Among observations with non-zero divergence, predicted JSD values are generally moderate and change little over time. In 2015-04 the conditional magnitude is around $0.31$--$0.32$ for all three pairs, and by 2022-10 it remains around $0.32$--$0.35$. The end-point comparison therefore suggests little overall growth in the conditional magnitude of affective divergence, even though non-zero divergence becomes more likely.

There are nonetheless some temporary hikes in conditional magnitude (CR-LL reaching $0.47$ in 2021-04, CR-MR reaching $0.55$ in 2017-10) whose timing is not directly explained by any specific political or climate-related event. However, the estimated marginal means and pairwise contrasts indicate that differences between pairs are usually small, with most credible intervals for the pair-wise ratios including $1.00$. These contrasts suggest at best episodic pair-specific differences in the magnitude of affective divergence.

The control variable indicates that pair-level activity is strongly related to conditional JSD. The coefficient for the log-transformed geometric mean of participation from both blocs in a pair per article is credibly negative ($-0.95$, $[-1.03,\,-0.88]$), equivalent to an approximately 61\% reduction in the odds of conditional divergence per unit increase in activity. Substantively, more viral articles tend to generate more similar valence distributions across blocs, implying that highly visible articles attract a larger share of neutral or affectively similar responses.

The conditional magnitude model provides a fairly good account of variation in non-zero JSD, with a Bayesian $R^2$ of $0.41$ $[0.38,\,0.43]$. Posterior predictive checks indicate that this model captures the observed mean, dispersion and range, although some extreme observations are less closely caught. Convergence diagnostics are again satisfactory, with all $\widehat{R}$ values effectively at $1.00$ and adequate effective sample sizes.

Affective divergence between ideological blocs is common and becomes more likely over time, particularly for pairs involving CR users. However, conditional on divergence being present, its magnitude remains moderate and only episodically differs across bloc pairs, suggesting that affective assimilation is present but not a strongly or consistently differentiated mechanism of polarisation under low media-party parallelism.

\subsection*{Ideological sorting}

\begin{figure}[ht]
    \centering
    \captionsetup{justification=centering,font=small}
        \includegraphics[width=0.99\textwidth, trim={0pt 0pt 0pt 0pt}, clip]{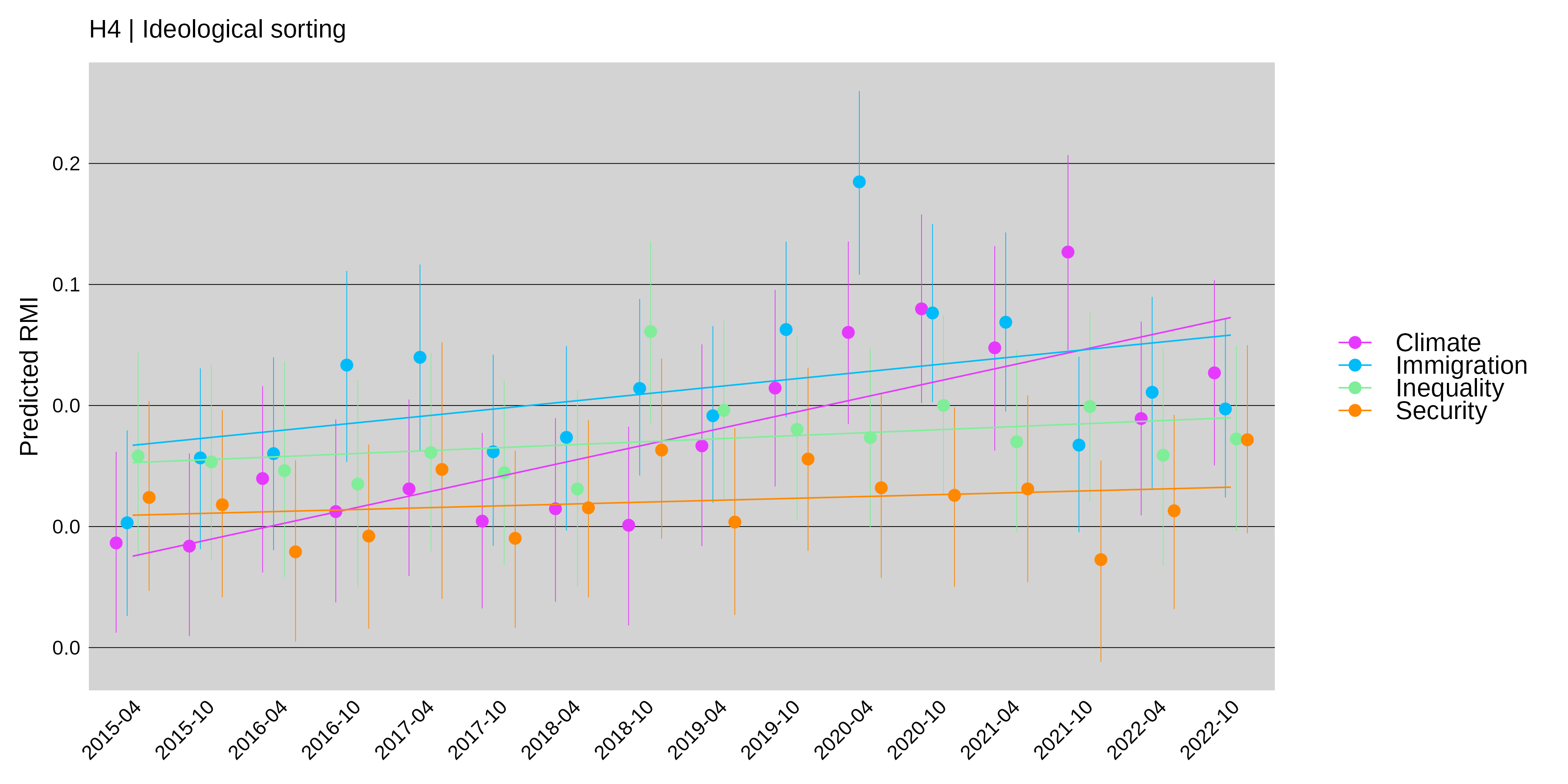}
        \caption{Predicted alignment, measured using reduced mutual information (RMI), between ideological blocs and issue-specific news-co-posting patterns over time; points denote posterior means and error bars indicate 95\% credible intervals; trend lines fitted separately to enhance interpretability.}
        \label{fig:ideological_sorting}
\end{figure}

Lastly, in Figure~\ref{fig:ideological_sorting}, we present the model-estimated RMI between the dynamic issue-specific news-co-posting communities and the ideological blocs which remain static over each four-year parliamentary cycle. Across all four issues, RMI remains modest, indicating that issue-specific co-posting communities only marginally align along the ideological blocs, although climate and immigration show clearer periods of alignment from 2019 onwards.


RMI remains close to zero across all four issues throughout the first parliamentary cycle, with all early credible intervals including zero and thus little evidence of systematic issue-ideology alignment at the outset. The estimates increase later in the series, but the timing and strength differ across issues. Climate shows the clearest late-period rise, remaining near zero until 2018-10 before increasing to a peak of $0.11$ $[0.07,\,0.15]$ in 2021-10, then declining to $0.04$--$0.06$ by 2022. This indicates a period of stronger ideological sorting from 2019-10 to 2021-10, coinciding with the 2019 parliamentary election, COVID-19 and major developments in climate politics including the IPCC reporting cycle; the 2022 decline is consistent with the Russian invasion of Ukraine temporarily reconfiguring issue-specific attention, though the model does not identify this mechanism directly. Immigration shows an earlier and sharper peak, rising to $0.14$ $[0.10,\,0.18]$ in 2020-04 (the highest estimate for any issue and time point) before declining to $0.03$--$0.06$ by 2021--2022, overlapping with the 2019 election and early COVID-19 period.

Inequality and security show weaker and less stable alignment. Inequality is mostly close to zero but reaches $0.08$ $[0.04,\,0.12]$ in 2018-10 and stays modestly positive ($\sim\!0.05$) in several later waves, suggesting some sorting but not the sustained convergence seen for climate or the sharp peak seen for immigration. Security is the weakest case, with estimates generally close to zero and most credible intervals including zero throughout, including during COVID-19 and after the Russian invasion of Ukraine.

The pairwise contrasts show that climate is not systematically more aligned than the other issues throughout the period. In several earlier waves it is instead lower than immigration or inequality (e.g. $-0.06$ to $-0.08$ relative to each in 2016-10--2018-10). This reverses after 2020, when climate and immigration become indistinguishable by 2020-10, both exceeding security by about $0.08$, and climate shows its clearest separation from all other issues in 2021-10 (exceeding immigration by $0.08$, inequality by $0.06$ and security by $0.13$). By 2022-10, however, the differences between climate and the other issues are again uncertain, with all contrasts close to zero.

The coefficient for the log-transformed size of the graph for which we compute the alignment is small and positive ($0.02$, $[0.00,\,0.04]$). This contributes little evidence that larger issue-specific graphs are associated with higher RMI after accounting for issue and time.

The model accounts for a substantial amount of variation in RMI, with a Bayesian $R^2$ of $0.70$ $[0.64,\,0.75]$. Posterior predictive checks indicate that the model reproduces the main distributional features of the observed RMI values well. Convergence and sampling diagnostics are again satisfactory, with $\widehat{R}=1.00$ and adequate effective sample sizes throughout.

Overall, what we observe is some movement towards ideological sorting in issue-specific news-co-posting, but this movement is uneven across issues and appears most clearly around specific political periods when polarisation appears to activate. The strongest convergence occurs around 2019--2021, especially for immigration and climate, while inequality shows a more modest and less sustained increase and security remains weakly aligned. Yet the modest magnitude of RMI throughout the series suggests that one certainly cannot reduce issue-specific news engagement to broader ideological divisions alone.

\section*{\large DISCUSSION}

We set out to explain the recent polarisation of climate politics in Finland, an unlikely setting for such polarisation given its low media-party parallelism, independent press and high trust in news among the public \cite{van_kempen_media-party_2007, newman_digital_2024}. Unlike much prior research trying to locate the main effect of media on climate politics at the publisher, we proposed four mechanisms operating downstream of publication and tested each against eight years of Finnish Twitter data, 2015--2023, spanning two parliamentary cycles: audience capture (H1), selective exposure (H2), affective assimilation (H3) and ideological sorting (H4). Each traces polarisation beyond publication itself, from outlet choice and article selection to valence expression and broader identity-driven sorting. Through a Bayesian regression approach, we tested each corresponding hypothesis in a previously understudied empirical setting.

First, our evidence speaks most directly to theories of media effects developed for systems characterised by high media-party parallelism. Prior work linking media structure to political polarisation often presumes that editorial alignment with party platforms drives the latter \cite{seymour-ure_political_1974, hallin_comparing_2004, martin_bias_2017}. Our \textit{strong support for H1} suggests that this framework does not directly generalise to consensual, low-parallelism democracies such as Finland. We found notable audience differentiation between mainstream and tabloid outlets, with CR (Conservative Right) users increasingly engaging with tabloid content relative to LL (Liberal Left) and MR (Moderate Right) users. This points to an audience-driven form of parallelism in which political differentiation emerges not from editorial position-taking by outlets but from media use. It resonates with \citeauthor{chadwick_tabloids_2018} (2018) whose media-as-resources perspective views news as a resource that audiences select and repurpose according to their own motivations and commitments. Our work extends this perspective by demonstrating that the elective affinity of tabloid news with particular forms of political engagement can also manifest as audience differentiation in a low-parallelism media system, creating an audience-capture-like feedback loop between differentiated audiences and editorial incentives \cite{jurg_alex_2025}.

Second, further evidence suggests that this differentiation does not end at the level of outlet choice. Within the same outlets, ideological blocs at times select more of the different than of the same articles. Yet the pattern is asymmetric, being most apparent in the CR-LL relationship, altogether providing \textit{partial support for H2}. Hence an implication for prior theoretical work on selective exposure \cite{gentzkow_ideological_2010, tyler_partisan_2022}: where one cannot assume audiences to differentiate simply by favouring ideologically pleasing outlets, polarisation can also arise through differentiated selection within the same outlet, within the same editorial line. In a low-parallelism context, within-outlet selective exposure therefore appears to constitute an additional layer of audience differentiation rather than a more general mechanism producing entirely separate information diets on the climate issue.

Third, evidence further down the chain from publication to reception yields \textit{moderate support for H3}. This hypothesis dealt with the valence with which users respond when they encounter the same climate-related news, as they presumably often do in a low-parallelism context. Interpreting affective divergence between ideological blocs as a tractable manifestation of affective assimilation \cite{han_between_2026, wilson_preferences_1989}, we found that non-zero divergence becomes somewhat more common over time, but its conditional magnitude stays moderate and the differences between ideological bloc pairs are episodic, albeit non-trivial. Put another way, users do not consistently transform identical climate-related articles into sharply opposing valence expressions according to ideological bloc. The evidence is therefore compatible with theories of motivated reasoning and biased assimilation which predict that prior expectations can well shape responses to the same piece of information \cite{lord_biased_2009, taber_motivated_2009}. Yet the affective expression of this process is less pronounced than the differentiation evident in outlet choice and, to a lesser extent, article selection.

Fourth, the model addressing our last hypothesis, effectively extending the analysis from individual behaviour to the structure emerging from broader identity-driven political organisation at the level of an entire political system \cite{baldassarri_partisans_2008, iyengar_affect_2012}, gave only \textit{weak support for H4}. We took ideological sorting as a process in which issue-specific communities emerging and evolving dynamically from news co-posting patterns increasingly reflect the more static ideological divisions. While some issue-specific communities indicate waves of greater alignment with the ideological blocs, particularly around 2019--2021, the absolute alignment remains modest throughout and does not indicate any clear correspondence across all four issues under consideration (immigration, inequality and security in addition to climate); nonetheless, the small absolute increase does not rule out ideological sorting from governing news engagement across the climate and immigration issues specifically, both of which exhibit an upward and largely parallel trend. Research looking at the Finnish political system around the 2019 election has also flagged the co-evolution of polarisation across these issues as resulting from partisan sorting \cite{chen_polarization_2021}. The steady, low alignment of security-specific communities with ideological blocs in turn ought to reflect the long-standing, albeit not unambiguous, consensus on foreign policy in Finland, especially after the 2022 Russian invasion of Ukraine \cite{xia_russian_2024}.

In conclusion, the four hypotheses point to an uneven accumulation of polarisation of climate politics across analytical levels, strongest at the media level and progressively weaker further down the chain towards system-level sorting. The Finnish case suggests that substantial political differentiation can emerge even where media-party parallelism is low. However, differentiation at one level does not necessarily translate into stronger polarisation at the next. The findings point to a multi-level process in which audience practices, content selection and affective expression interact within a contemporary hybrid media system, while broader ideological sorting guiding news engagement remains fairly weak, perhaps surprisingly.

From a methodological perspective, our findings underline the usefulness of modelling polarisation as a multi-dimensional process. Ideological differentiation in outlet choice, within-outlet article selection, affective expression and system-level sorting do not need evolve synchronously, and collapsing them into a common measure would blur the differences which our four hypotheses were meant to identify. Polarisation is an emergent macro-level phenomenon and as such manifests in many ways, including for instance opinion, affect and lifestyle \cite{iyengar_affect_2012, dellaposta_pluralistic_2020, brown_measurement_2021}. Yet several aspects of our operationalisations warrant caution. For example H2 relies on AEIs computed from the largest component of each outlet-time projection, meaning that users who do not co-post any article in the same window contribute nothing to the index. If such isolates are unevenly distributed across ideological blocs, this may bias estimates of within-outlet segregation. The comparatively modest explanatory power of the H2 model, together with incomplete fit in the H3 and H4 models, also points to unobserved heterogeneity. Outlet-, article- and user-level characteristics likely contribute to which particular articles travel within a bloc, including framing, timing, author and the overall virality of the article, that our data cannot directly capture.

H3 is further limited by its reliance on a sentiment classifier applied to short posts which often contain little context. Future analyses should complement valence classification with explicit measures of stance, framing and argumentative structure, including through recent approaches based on automated discourse network analysis \cite{angst_automated_2025}. Even where users belonging to different blocs respond with similar valence, the underlying framing, especially in climate politics, might in fact diverge considerably.

The weak evidence for H4 also carries a methodological implication. Reduced mutual information by \citeauthor{newman_improved_2020} (2020) is appropriate for comparing partitions with different cluster sizes and numbers of communities, but no single measure to date exhausts the concept of structural correspondence. State-of-the-art information-theoretic measures are also rather challenging to interpret because the amount of additional shared information required to increase the value depends on the underlying partition structure, meaning that equal numerical differences do not necessarily represent equivalent changes in correspondence. We circumvented this difficulty by focusing on changes in RMI over time.

As regards the substantive implications of our study, the pronounced role of CR users across our analyses also situates the Finnish case within a wider transformation of climate politics. Recent research increasingly identifies radical-right and right-wing populist actors as important drivers of climate conflict, opposition and political mobilisation across a range of national contexts \cite{dickson_going_2025, bosetti_green_2025}. Our findings do not demonstrate that Finnish CR users constitute a homogeneous radical-right constituency, nor that tabloid engagement causes climate scepticism or opposition. They nevertheless suggest that the increasing politicisation of climate news might be asymmetric, concentrating disproportionately on the right-ward edge of the ideological space. If public disagreement results increasingly from audience differentiation instead of exposure to overtly partisan information, formal editorial neutrality might do little to hinder or revert polarisation as such.

There are also some more general limitations. First, our analysis is observational and cannot establish the direction of causality between audience behaviour and media strategy. Changing audience composition is bound to encourage adaptation but changing editorial or distribution strategies might also reshape audience composition. Second, the three-bloc solution, chosen to maintain comparability across the consecutive parliamentary cycles, necessarily simplifies a more continuous and potentially heterogeneous ideological space. Third, although the sentiment classifier performs reasonably well on held-out data, it was not trained on an identical domain and was deliberately conservative in assigning non-neutral labels, meaning that the prevalence of affective engagement in our data could be lower than in reality. Twitter users in the study period were also not representative of the Finnish population, yet Twitter itself changed substantially in ownership, governance and algorithm after the study period, limiting direct generalisation to the contemporary platform environment.

Comparative replication across other democratic corporatist and polarised pluralist media systems would further establish whether the combination observed here, namely strong outlet-level differentiation, more limited within-outlet selective exposure, modest affective assimilation and weak ideological sorting, is characteristic of low-parallelism systems beyond Finland. Extending the analysis across longer periods and additional salient issues could clarify whether the relatively weak correspondence between issue-specific communities and ideological blocs reflects the continuing autonomy of climate politics or an earlier stage in a longer process of ideological bundling. Taking on the task of unravelling political polarisation within and across its various dimensions, across the increasingly fragmented hybrid media system, also warrants urgent attention to advance our understanding of the factors behind the current impasse in global climate action \cite{judge_environmental_2023}.



\section*{\large ACKNOWLEDGEMENTS}

We thank Jussi-Veikka Hynynen, Ali Salloum, Mikko Kivelä, Tuomas Ylä-Anttila and Jeffrey T. Hancock for their useful inputs on our work. Our work was made possible with funding from the Helsingin Sanomat Foundation $[20210021]$, the Finnish Foundation for Economic Education $[22-12164]$ and the Strategic Research Council of the Research Council of Finland $[352561]$. 

\bibliographystyle{apacite}
\bibliography{bibliography}

\newpage
\section*{\large APPENDIX}
\label{sec:appendix}

\subsection*{Inferring affective valence of posts}

To estimate the affective valence accompanying a post containing a hyperlink to a news article, we trained a sentiment classifier for the Finnish language with negative, neutral and positive outcome classes. The model relied on the \textit{FinnSentiment} corpus containing 27,000 sentiment-annotated comments in Finnish from the online portal \textit{Suomi24} \cite{linden_finnsentiment_2023}. The data were randomly shuffled and partitioned into training, validation and test sets, with 90\% of observations used for training and 5\% each for validation and testing.

The classifier in turn used a pre-trained Finnish MegatronBERT model as a foundation \cite{kanerva_finnish_2021}, which at 1.3 billion parameters was one of the largest natural language understanding models for Finnish at the time of analysis. The input text was tokenised using the model-specific tokeniser with a maximum sequence length of 128 tokens; longer texts were truncated and shorter ones padded to this length. We fine-tuned the model for the downstream classification task using the \textit{AdamW} optimiser with a learning rate of \(2 \times 10^{-5}\), batch size of 16, weight decay of 0.1 and no warm up. Hyperparameters were selected through a grid search over batch size, learning rate and warm-up ratio, where 12 configurations were evaluated over four training epochs each and the best-performing checkpoint retained. The final model was trained for one epoch using the selected configuration.

Because no publicly available Finnish sentiment datasets existed for the target domain of Twitter, the model was trained on data drawn from a related but distinct domain. Although both corpora consist of short-form social media text, differences in structure and usage introduce domain shift. In particular, posts frequently contained hashtags, user mentions and embedded hyperlinks, and often included quoted headlines or excerpts from linked articles. These features were largely absent from the training data and could have affected classification performance. Moreover, posts in our sample tended to react to external content rather than participate in conversational threads, and the distribution and expression of valence appeared more restrained than in \textit{Suomi24} discussions. To mitigate these issues, several pre-processing steps were applied prior to classification. Hyperlinks were removed, sequences of multiple hashtags were omitted and single hashtags were normalised by removing the hash symbol. In addition, recurring n-grams associated with shared hyperlinks were identified and removed to exclude article headlines and quotations.

Further adjustments were made at the prediction stage to address the tendency of the model to overpredict the neutral class. Specifically, posts assigned a neutral label with low confidence were re-evaluated by comparing the predicted probabilities of the negative and positive classes. If the model showed a clear preference for one of these classes, the label was reassigned accordingly, while ambiguous cases were retained as neutral. The probability thresholds governing this adjustment were optimised on a manually annotated validation set and validated on a held-out sample. The resulting model achieved an accuracy of $0.88$ on the FinnSentiment test set. When applied to domain-specific data, performance improved substantially after the prediction adjustment procedure, indicating that the post-processing step helped correct for domain-induced bias in class probabilities.

\begin{figure}[H]
    \centering
    \captionsetup{justification=centering,font=small}
        \includegraphics[width=0.99\textwidth, trim={0pt 0pt 0pt 0pt}, clip]{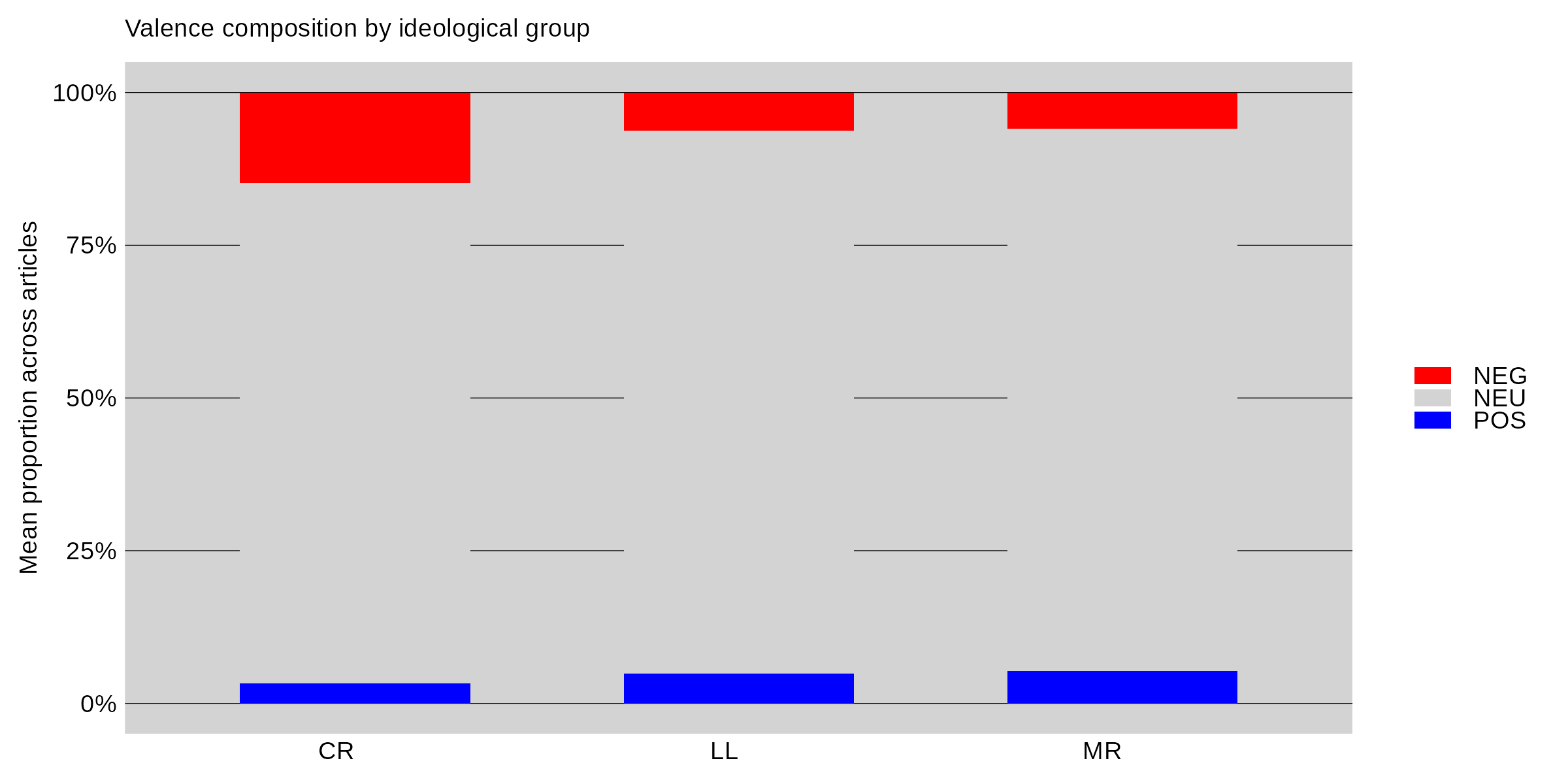}
        \caption{Mean article-level valence composition in climate-related posts by ideological bloc.}
        \label{fig:valence}
\end{figure}

Figure~\ref{fig:valence} presents the distribution of valence across all climate-related posts in our data by ideological bloc, spanning the entire study period. Posts with non-neutral valence account for 11\% of all posts, while the share of negative valence is slightly more pronounced among conservative-right (CR) users. This baseline distribution helps us to understand the relatively low JSDs across articles, given that non-neutral reactions are fairly uncommon on Finnish Twitter in the study period.

\subsection*{Inferring ideological leaning of users}

To assign each user a political stance, we assigned users to groups according to their broader retweeting behaviour with reference to Finnish political parties. We began by constructing an undirected and unweighted unipartite graph of all retweets among the users containing the name of at least one of the major political parties in Finland. Collecting these data entailed passing the ``parties'' query, in August 2023, to Twitter's Academic programming interface (Table~\ref{tab:keyword_filters}).

After parsing and cleaning these data, we created a separate graph for each parliamentary cycle (Sipilä cabinet: 2015/04/19-2019/04/14, Marin cabinet: 2019/04/15--2023/04/02). We removed all users who only had a single connection, and extracted the largest component of the graph. Following \citeauthor{wagner_breaking_2025} (2025), we then inferred the assortative modular structure of the graph (i.e., groups of vertices that are more connected to one another than to vertices of other groups) by combining two graph clustering techniques: modularity optimisation and stochastic block modelling, both of which alone suffer from certain deficiencies, regarding for instance model selection or interpretability, respectively.

We used the Leiden algorithm by \citeauthor{traag_louvain_2019} (2019) to optimise modularity, a much-used measure of the quality of the partition, for a range of resolution parameters (from 0.00 to 1.50 at 0.01 intervals). Resolution adjusts for the size of modules and its value is crucial for uncovering any meaningful modular structure, but the selection of which is often entirely arbitrary as such. As modularity values are not comparable across different resolutions or might arise from random fluctuations, we performed model selection by passing each ``proposal'' from the Leiden algorithm to a degree-corrected planted partition model to determine the description length of the solution under the data \cite{zhang_statistical_2020}. By selecting the model according to minimum description length (i.e. the minimum amount of information that is necessary to describe the data), we gained statistical evidence for the existence of a modular structure (against alternative solutions, including a single-module solution) and effectively avoided overfitting, or underfitting thereof, the model (i.e., finding modules where there really are not any or the other way around). As the Leiden algorithm is stochastic, for each resolution parameter, we used the highest modularity-yielding partition over 20 runs.

\begin{figure}[H]
    \centering
    \captionsetup{justification=centering,font=small}
        \includegraphics[width=0.495\textwidth, trim={0pt 0pt 0pt 0pt}, clip]{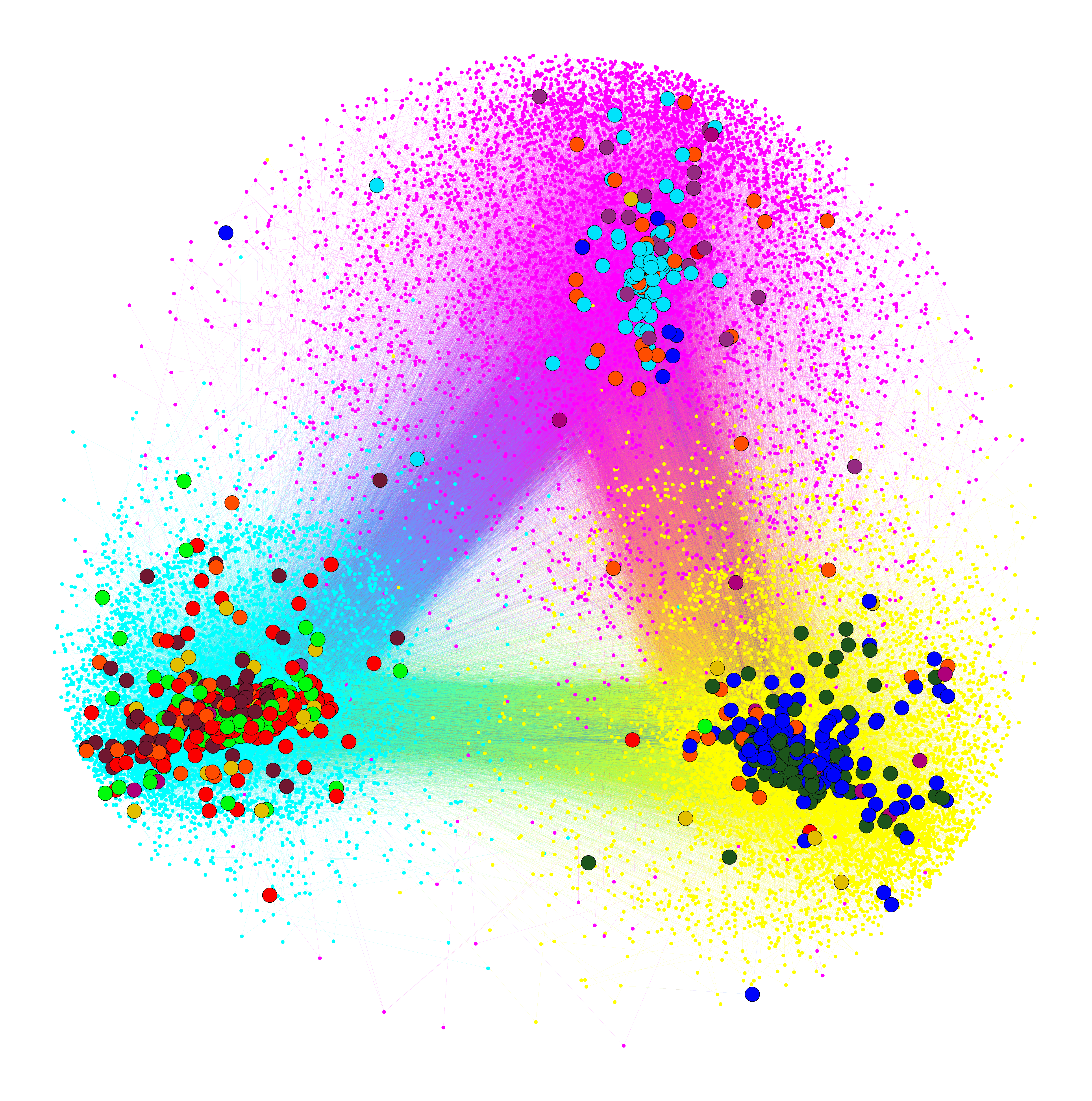}
        \includegraphics[width=0.495\textwidth, trim={0pt 0pt 0pt 0pt}, clip]{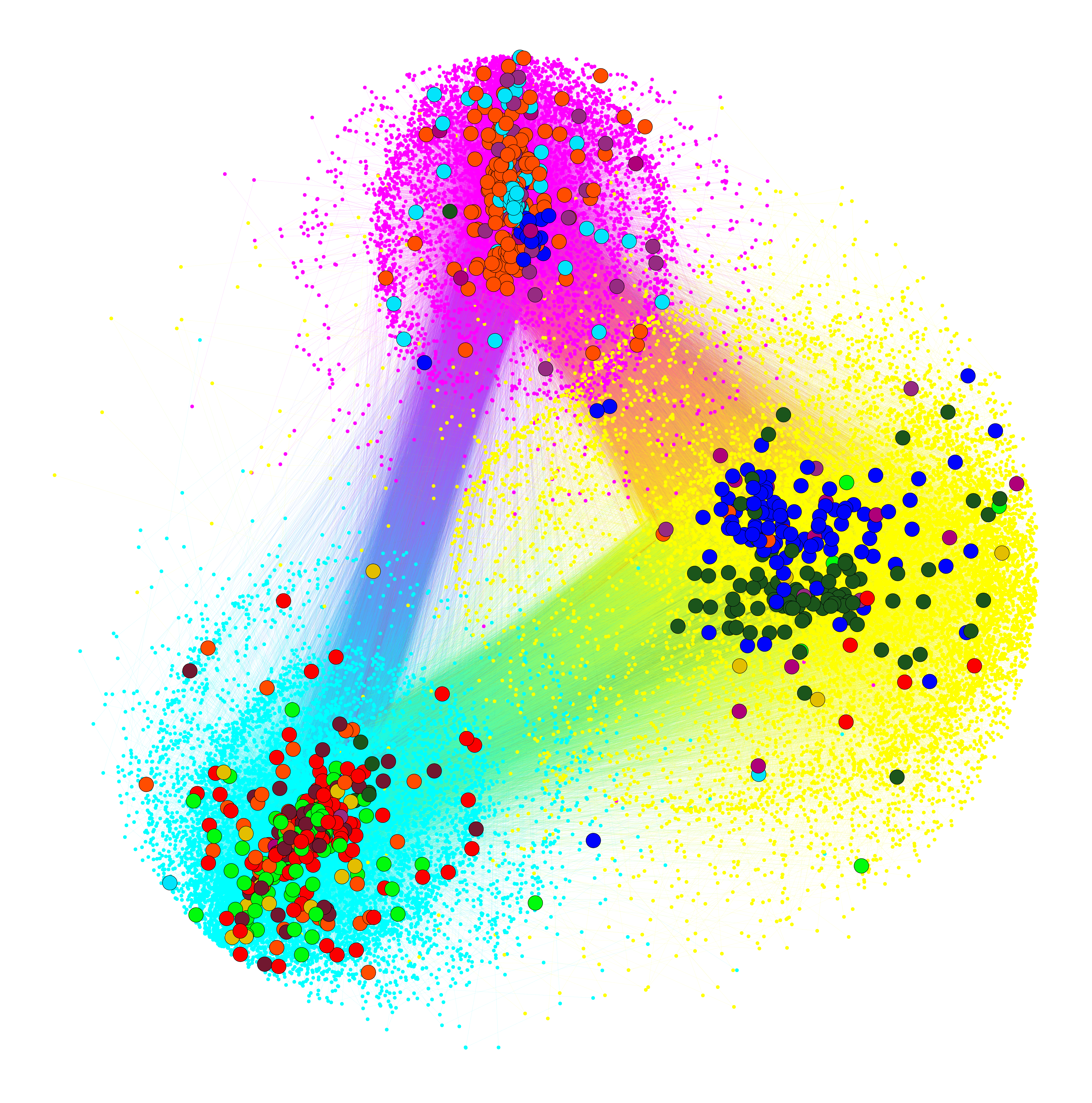}
        \caption{Stance estimation of Twitter users in Finland via cluster assignments in the political party repost graph in 2015--19 (left) and 2019--23 (right), highlighting the accounts of candidates who ran in the 2019 and 2023 parliamentary elections; \textcolor[HTML]{ff00ff}{{\Huge\textbullet}} Conservative Right, \textcolor[HTML]{00ffff}{{\Huge\textbullet}} Liberal Left, \textcolor[HTML]{ffff00}{{\Huge\textbullet}} Moderate Right, \textcolor[HTML]{0307fc}{{\textbullet}} National Coalition Party, \textcolor[HTML]{03e3fc}{{\textbullet}} Finns Party, \textcolor[HTML]{1d5720}{{\textbullet}} Centre Party, \textcolor[HTML]{03fc0f}{{\textbullet}} Green League, \textcolor[HTML]{fc0303}{{\textbullet}} Social Democratic Party, \textcolor[HTML]{751a34}{{\textbullet}} Left Alliance, \textcolor[HTML]{e3be02}{{\textbullet}} Swedish Party of Finland, \textcolor[HTML]{942e83}{{\textbullet}} Christian Democratic Party, \textcolor[HTML]{af0078}{{\textbullet}} Movement Now, \textcolor[HTML]{ff4d00}{{\textbullet}} Other parties without representation in the Finnish parliament.}
        \label{fig:ideology_inference}
\end{figure}

The shortest description lengths for the two consecutive graphs suggested five and three modules of users, respectively. To retain comparability across cycles, we refined the search by capping the number of clusters to three. We also validated these clusters by mapping the assignments of known Twitter accounts of election candidates running either in the 2019 or 2023 parliamentary elections. As shown in Figure~\ref{fig:ideology_inference}, the resulting counts of candidates representing different parties per module were used for inspecting, labelling and aligning the blocs. These ideological blocs align well with what is known of the long-term stances of Finnish political parties and therefore laid the basis for extending the module assignment as representative of the ideological leaning of a user \cite{isotalo_polarisoituuko_2020}, whether conservative right (CR), liberal left (LL) or moderate right (MR). The benefit of performing this task separately for each four-year parliamentary cycle is that we allow for users to update their stance from one cycle to the other, while facilitating interpretation of our main analyses by fixing a key attribute of the target population.

\begin{figure}[H]
    \centering
    \captionsetup{justification=centering,font=small}
        \includegraphics[width=0.99\textwidth, trim={0pt 0pt 0pt 0pt}, clip]{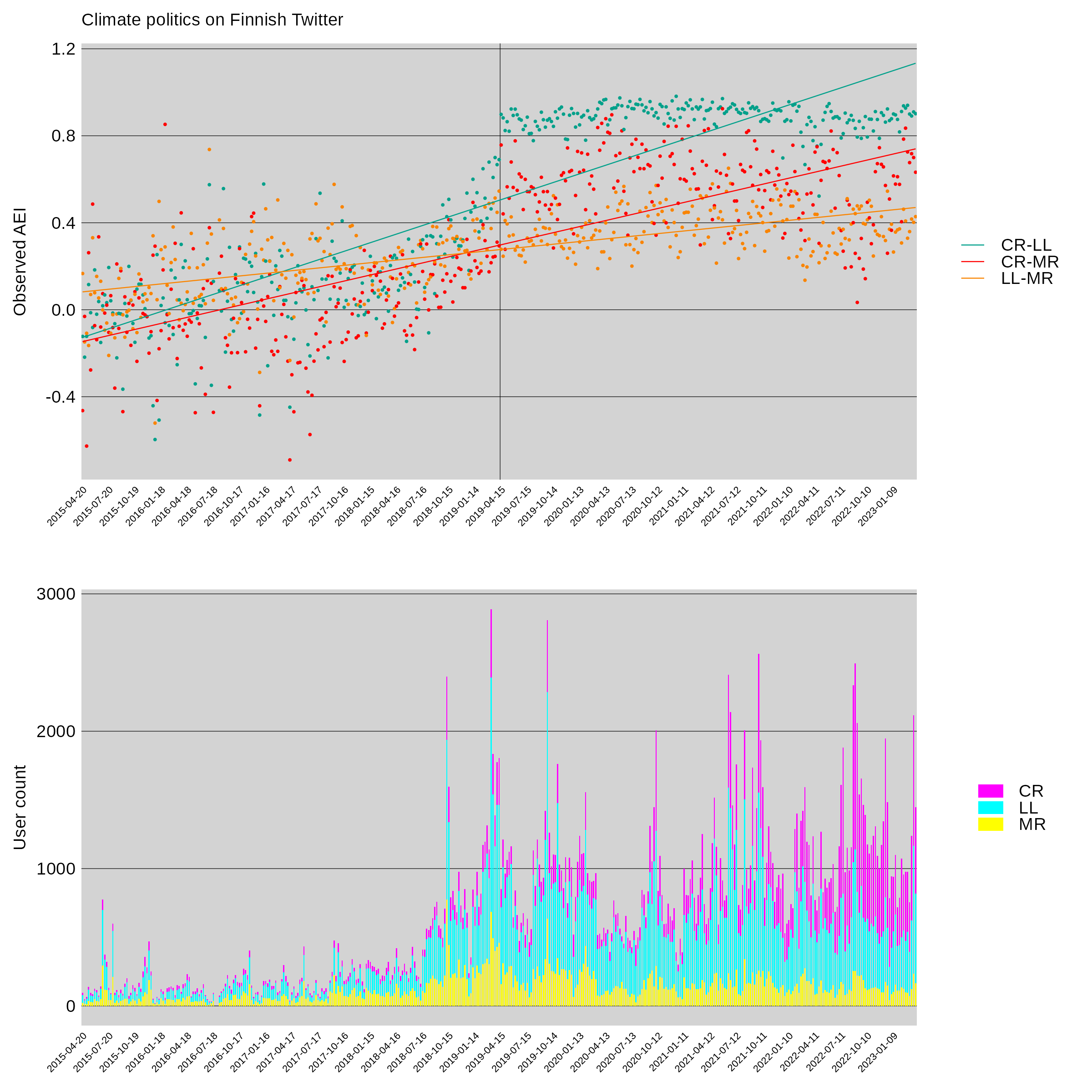}
        \caption{Polarisation of and participation in climate politics on Finnish Twitter from April 2015 to April 2023; black vertical line in the upper panel denotes the change in parliamentary cycle.}.
        \label{fig:validation}
\end{figure}

As an auxiliary validation of the bloc assignments, we computed a weekly polarisation score based on users' climate-related retweeting behaviour between each bloc pair over the study period (Figure~\ref{fig:validation}). To estimate polarisation, we again used the AEI index by \citeauthor{salloum_separating_2022} (2022), but unlike in our H2 operationalisation, we simply log-transformed, with a constant, the natural edge weights corresponding to retweet counts between a pair of users. A value of one here implies that there are not any connections between blocs, a value of zero means that the weight within and between them is equal and the value of minus one ought to indicate that all connections occur between blocs, yielding an easily interpretable operationalisation of retweet polarisation \cite{malkamaki_connective_2026}. Figure~\ref{fig:validation} also shows the corresponding weekly retweeting activity of these users. The median weekly share of users participating in climate politics but lacking a bloc assignment is 18\%, whereas the values vary between 4\% and 76\% over the study period. These users were omitted from all analyses. Polarisation between CR and LL, who also populate the discussion on the topic, increases and stays high since 2018. The polarisation between LL and MR is notably lower throughout the period, while between CR and MR there is a sudden yet dramatic drop right after the Russian invasion of Ukraine in 2022, corroborating a temporary depolarisation process as recently shown by \citeauthor{xia_russian_2024} (2024). Overall, these trends match our understanding of the political shifts and allegiances over the study period, providing further validation to the robustness of our ideology inference approach.

In our operationalisation of ideological sorting (H4), we use a similar approach to identify the assortative structures from users' issue-specific news-posting per each three-month time slice. However we cap the number of possible communities between two and six, always choosing the partition yielding the shortest description length of a degree-corrected stochastic block model. Unlike when inferring the ideological blocs, here we do not simply omit single-connection users and treat the resulting edges as unweighted but include the cosine-normalised edge weights of the relevant bipartite projection as a covariate of an exponential form \cite{peixoto_nonparametric_2018}.

\subsection*{Keyword filters}

\begin{table}[H]
    \centering \small
    \caption{Partial match keyword filtering for climate, immigration, inequality and security data; parties data rely entirely on the original query to Twitter's academic programming interface; access to all original queries via [placeholder].}
    \begin{tabularx}{\linewidth}{@{} lX @{}}
        \toprule
        \textbf{Climate} & extinction rebellion, vihreä siirtymä, vihreän siirtymä, bensakapina, cleantech, elokapina, extinctionrebellion, fossiili, geoterm, greta, hiilidioksidi, hiilinielu, ilmasto, ipcc, irtikytkentä, kasvihuon, koululak, lentolak, lentovero, lihavero, lulucf, metaani, nytonpakko, perjantaitgretankanssa, päästö, reilusiirtymä, reilu siirtymä, reilun siirtym, thunberg, tietull, turbiini, turpee, turve, tuuliturbiini, tuulivoima, uniper, päästöjä, ydinvoima, ydinreaktor, ydinjät, pienreaktor, ydinturvallisuu, atomivoima, atomireaktor, atomiturvallisuu, olkiluo, ol3, co2, fennovoima, hanhikiv, aurinkosähkö, vihreään siirtymä, aurinkovoima, tuulienergi, aurinkoenergi, ydinenergi \\
        \textbf{Immigration} & kansainvälisen suojel, kansainvälistä suojel, kansanvälinen suojel, kansanväliseen suojel, laiton maahantul, laittoman maahantul, maassa maan tavalla, rajat kiinni, tilapäinen suojelu, tilapäistä suojelu, työperäinen maahanmuut, työperäisen maahanmuut, työperäistä maahanmuut, vapaa liikkuvuus, vapaaehtoinen paluu, ankkurilapsi, haittamaahanmuut, ihmiskauppa, integr, kaksoiskansalai, kotouttami, lähtömaa maahanmuut, maahantul, maahantunkeutuj, mamu, matu, migri, muuttoliik, (derogatory term for a person of colour), oleskelulu, oleskeluoikeu, oleskelustatu, ongelmalähiö, pakkopalauttami, pakolai, paperiton, paperittomi, perheenyhdistämi, rajatkiinni, rajaturvallisuu, refugee, siirtolai, säilöönotto, tphakija, turvapaikanhakija, turvapaikanhakijakiintiö, turvapaikkakiintiö, kaksoiskansalai, kiintiöpakolai \\
        \textbf{Inequality} & vero, tuloero, varallisuusero, tasa-arvo, eriarvo, tuloluok, tulonsaaj, minimipalkk, vähimmäistulo, vähimmäispalk, gini, listaamattom, rikka, tulonmuun, tulojen muun, pääoma \\
        \textbf{Security} & nato, hävittäj, hx-hank, hxhanke, hornet, gripen, f35, f-35, krim, hybridiuhk, hybridivaikut, informaatiovaikut, ottawan, maamiin, puolustu, tp-utva, tputva, turvallisuuspolit, turpo, palkka-armeija, sotilasliit, natsi, ukropp \\
        \textbf{Parties} & (kokoomus OR kokoomuslainen OR kokoomuslaisen OR kokoomuslaista OR kokoomuslaiset OR kokoomuslaisten OR kokkari OR kokkarit OR kokkareiden OR perussuomalainen OR perussuomalaiset OR perussuomalaisen OR perussuomalaisten OR persu OR persut OR persujen OR ps OR sdp OR sosiaalidemokraatit OR sosiaalidemokraattisen OR sosiaalidemokraattinen OR sosiaalidemokraattien OR demari OR demarit OR demarien OR demareiden OR keskustapuolue OR keskustapuolueen OR keskusta OR keskustan OR keskustalaisen OR keskustalaiset OR keskustalaisten OR kepu OR kepulainen OR kepulaiset OR kepulaisten OR vihreät OR vihreiden OR vasemmistoliitto OR vasemmistoliiton OR vasemmistoliitosta OR vasemmisto OR vasemmiston OR vasemmistolaiset OR vasemmistolainen OR vassari OR vassarit OR vassareiden OR vihervasemmisto OR vihervasemmiston OR vihervasemmistolaiset OR vihervasemmistolainen OR vihervassari OR vihervassarit OR vihervassareiden OR punavihreä OR punavihreät OR punavihreän OR punavihreiden) lang:fi \\
        \bottomrule
    \end{tabularx}
    \label{tab:keyword_filters}
\end{table}

\subsection*{Outlet categories}

\begin{table}[H]
    \centering \small
    \caption{Post-harmonisation domains of relevant outlets by category; partisan category includes \textit{mt.fi} and \textit{alfatvuutiset.fi} which do not formally represent any political party but respectively align closely with Centre Party's and Christian Democrats' interests and organs, as well as \textit{vihrealanka.fi} that was an official organ of the Green League until its controversial closure and \textit{de facto} replacement with \textit{verdelehti.fi} in 2019, and later reinvigoration by a faction within the party.}
    \begin{tabularx}{\linewidth}{@{} lX @{}}
        \toprule
        \textbf{Financial} & kl.fi, talouselama.fi, tekniikkatalous.fi \\
        \textbf{Mainstream} & yle.fi, mtvuutiset.fi, hs.fi \\
        \textbf{Miscellaneous} & keski-uusimaa.fi, aamuposti.fi, hbl.fi, esaimaa.fi, kouvolansanomat.fi, kymensanomat.fi, lansi-savo.fi, sss.fi, uusimaa.fi, vasabladet.fi, uutiset.live, uusisuomi.fi, suomenkuvalehti.fi, maailmankuvalehti.fi, ylioppilaslehti.fi, viikkosavo.fi, kalajokilaakso.fi, lansi-uusimaa.fi, lappeenrannanuutiset.fi, lansivayla.fi, raahenseutu.fi, kurikka-lehti.fi, helsinginuutiset.fi, seinajokinen.fi, kaupunkilehti.fi, image.fi, versuslehti.fi, mustread.fi, muutoslehti.fi, suomenluonto.fi, audiomedia.fi, kirkkojakaupunki.fi, aamunkoitto.fi, tivi.fi, tekniikanmaailma.fi, seura.fi, eurojatalous.fi, porssiuutiset.fi, marmai.fi, menaiset.fi, seiska.fi \\
        \textbf{Fringe} & tokentube.net, positv.fi, mvlehti.net, partisaani.com, maaseutumedia.fi, oikeamedia.com, kansalainen.fi, nykysuomi.com, magneettimedia.com \\
        \textbf{Regional} & aamulehti.fi, kaleva.fi, ksml.fi, ts.fi, savonsanomat.fi, karjalainen.fi, lapinkansa.fi, sk.fi, ilkkapohjalainen.fi, kainuunsanomat.fi, keskipohjanmaa.fi, ess.fi \\
        \textbf{Partisan} & mt.fi, alfatvuutiset.fi, verdelehti.fi, vihrealanka.fi, demokraatti.fi, kdlehti.fi, ku.fi, tiedonantaja.fi, suomenmaa.fi, verkkouutiset.fi, suomenuutiset.fi \\
        \textbf{Tabloid} & il.fi, is.fi \\
        \bottomrule
    \end{tabularx}
    \label{tab:outlet_categories}
\end{table}

\end{document}